\documentstyle[11pt,aaspp4,flushrt]{article}

\def\lea{\mathrel{<\kern-1.0em\lower0.9ex\hbox{$\sim$}}}
\def\gea{\mathrel{>\kern-1.0em\lower0.9ex\hbox{$\sim$}}}
\newcommand{\dbv}{$\Delta ({\rm B} - {\rm V})$}
\newcommand{\dv}{\hbox {$\Delta \rm V$}}
\newcommand{\feh}{\hbox{$ [{\rm Fe}/{\rm H}]$}}
\newcommand{\mvrr}{\hbox {$\rm M_v(RR)$}}
\newcommand{\ea}{{\it et al.}}

\slugcomment{Invited Review to appear in PASP}

\lefthead{Sarajedini, Chaboyer, & Demarque}
\righthead{Relative Ages of Globular Clusters}

\begin{document}

\title{The Relative Ages of Galactic Globular Clusters}

\author{Ata Sarajedini\altaffilmark{1,}\altaffilmark{2}}
\affil{Kitt Peak National Observatory, National Optical Astronomy
Observatories\altaffilmark{3},\\ P. O. Box 26732, Tucson, AZ  85726\\
{\it ata@stars.sfsu.edu}}

\author{Brian Chaboyer\altaffilmark{1}}
\affil{Steward Observatory, University of Arizona, Tucson, AZ 85721\\
{\it chaboyer@as.arizona.edu}}

\and

\author{Pierre Demarque}
\affil{Department of Astronomy and Center for Solar and Space Research,\\ 
Yale University, Box 208101, New Haven, CT 06520-8101\\
{\it demarque@astro.yale.edu}}

\altaffiltext{1}{Hubble Fellow}

\altaffiltext{2}{Current address is Department of Physics and Astronomy,
San Francisco State University, 1600 Holloway Avenue, San Francisco, CA
94132}

\altaffiltext{3}{NOAO is operated
by the Association of Universities for Research in Astronomy, Inc., under
cooperative agreement with the National Science Foundation.}

\begin{abstract}
We present a review of the present state of knowledge regarding the
relative ages of Galactic globular clusters. First, we discuss the
relevant galaxy formation models and describe the detailed predictions
they make with respect to the formation timescale and chemical
evolution of the globular clusters. Next, the techniques used to
estimate globular cluster ages are described and evaluated with
particular emphasis on the advantages and disadvantages of each method.
With these techniques as a foundation, we present arguments in favor of
the following assertions: 1) The age of a globular cluster is the
likeliest candidate to be the global second parameter, which along with
metal abundance, controls the morphology of the horizonal branch. 2) A
total age range of as much as $\sim$5 Gyr exists among the bulk of the Galactic
globulars. 3) There is a significant relation between age and
metallicity among the Galactic globular clusters if the slope of the
\mvrr-\feh\ relation is less than $\sim$0.23. These conclusions along
with other supporting evidence favor a formation scenario in which
the inner regions of the Galactic halo collapsed in a monotonic fashion
over a short time period much less than 1 Gyr. In contrast, the outer
regions of the halo fragmented and collapsed in a chaotic manner
over several Gyrs. 

\end{abstract}

\section{Introduction}

The relative ages of the Galactic globular clusters are an important signpost
on the path to understanding the formation of the Milky Way. However, 
what they tell us about the earliest epochs of the Galaxy has been 
interpreted in a 
number of different ways by various investigators. This is compounded by
the fact that we have age information for only one-third of the approximately
150 known globular clusters in the Milky Way. The most recent review 
concerning the relative ages of Galactic globular clusters was presented 
by Stetson et al. (\cite{stvabo1996}, hereafter SVB). In addition, 
there has been a flurry
of papers written on this topic over the past year. As a result, we felt
it appropriate to write another review updating and clarifying some
of the points made by SVB and others.

To motivate the study of relative globular cluster ages, we first
present a summary of the two competing models for the formation of the
Galaxy. The next section describes the various techniques 
used to estimate relative ages.
We show how these ages along with the corresponding abundances of the 
Galactic globulars can
provide insight into which of these models is most relevant to the
early formation history of the Milky Way. We then attempt to synthesize 
a formation scenario that is consistent with the observational data.

\section{Milky Way Formation Models}

It is generally believed that the Milky Way began as a single large 
gas cloud with perhaps some angular
momentum. Once it decoupled from the overall Hubble flow, this gas cloud
began to collapse and thus form the Galactic halo and disk that we know
today. Precisely how this collapse progressed is an issue of some debate
at the moment. Models for the formation of the Milky Way's halo can be 
classified into two categories: those that predict a rapid collapse of 
the Galactic halo ($<<$1 Gyr) and those that predict a more gradual 
and more chaotic collapse (several Gyr). 

The principal representative of the rapid collapse picture is the
work of Eggen et al. (\cite{eglysa1962}, hereafter ELS). They studied a set 
of 221 solar neighborhood stars with a range of metallicities. ELS
observed that lower metallicity stars, inferred to be older, are on
more eccentric orbits than those with higher metallicity. In addition,
ELS found that lower metallicity stars seemed to have been formed
at a range of heights above the galactic plane whereas those with
higher metallicity were formed preferentially near the galactic plane.
When interpreted in the context of their model for the Galaxy, these
findings lead ELS to conclude that the Milky Way began as a spherical
rotating gas cloud with a radius of at least 100 kpc. This cloud
collapsed on a free-fall timescale of a few x $10^8$ years and spun up
while the first generation of stars and globular clusters formed. These
earliest stars were relatively metal-poor and on highly eccentric
orbits caused by the rapidly changing gravitational potential of the
gas cloud. ELS pointed out that the short timescale of this collapse
explains why all of the globular clusters for which accurate photometry
was available - M3, M5, M13, M15, and M92 - are so nearly of the same age.
One caveat that is immediately obvious from the work of ELS is that
their kinematical data only sampled the halo out to a maximum of
$\approx$10 kpc above the galactic plane. As a result, their conclusions
are necessarily only applicable to this portion of the halo. 

The gradual collapse scenario has been principally advocated by 
Searle \& Zinn (\cite{sezi1978}, hereafter SZ). However, before describing the
work of SZ, it is beneficial to discuss the developments between the time
of ELS and SZ. In particular, Sandage \& Wildey (\cite{sawi1967}, see also
van den Bergh \cite{van1967}) noticed that M3, M13, and NGC 7006 have very
similar metallicities but quite different HB morphologies - the HB of M13 is
completely blueward of the RR Lyrae instability strip, that of NGC 7006 is
almost completely redward, and M3 has both a blue and red HB. They knew that
the morphology of the HB is determined primarily by metal abundance
(the first parameter) based on the theoretical work of 
Faulkner (\cite{fau1966})
and Faulkner \& Iben (\cite{faib1966}), i.e. high metallicity leads to a red HB
morphology and vice versa. However, here were three clusters with the
same metallicity but vastly different HB morphologies. Thus, they
concluded that a second parameter must be at work affecting the appearance
of the HB, giving rise to the `Second Parameter Effect.' SZ were searching
for this second parameter. They analyzed the metallicites of globular 
clusters as a function of their Galactocentric distances. From this,
they concluded that outside of $\approx$8 kpc, the abundance distribution
of globular clusters is independent of galactocentric distance. This is
in contrast to the predictions of the ELS model which would predict a
dependence of abundance on radial distance. Next, SZ constructed the
[Fe/H] : HB-type diagram for globulars in different ranges of Galactocentric
distance. From this they noticed that the clusters of the inner halo
($R_{GC} < 8$ kpc) are not influenced by the second parameter whereas
those outside of 8 kpc are so influenced. They asked the question:
which of the many second parameter candidates can vary with Galactocentric
distance in the most plausible manner? They argued that the cluster age
is the most likely candidate to be the second parameter, implying a range
of ages among the globular clusters outside 8 kpc. As a result, the
following picture of halo formation emerges. The Galaxy began as a 
spherical gas cloud. The inner portions (inside 8 kpc) collapsed
quickly ($<$ 1 Gyr), but the outer parts fragmented in a chaotic fashion.
These fragments were on eccentric orbits and sometimes collided with each
other losing their angular momentum. After a few Gyr, the gaseous components
of these fragments merged with
the inner Galaxy while the stars and globular clusters formed in these 
fragments retain the orbits they possessed at the time of formation.

These then are the two basic formation scenarios of the Milky Way. 
An important part of galaxy formation models 
relates the dynamics of the collapsing gas cloud to the history of 
metal enrichment in the forming galaxy  (see e.g. Larson \cite{lar1975}; 
Tinsley \cite{tin1975}; Ostriker \& Thuan \cite{osth1975}; 
Hartwick \cite{har1976}).  These are 
relevant in understanding the formation of the Galactic halo and the 
age-metallicity relation (Larson \cite{lar1990}). As such, each formation model
makes specific testable predictions. In particular, the ELS model predicts
a small ($<<$ 1 Gyr) age spread among the Galactic globular clusters and
a strict one-to-one correspondence between the age of a cluster and
its metallicity; there is no implicit prediction regarding the nature
of the second paramter. The SZ model predicts an age range of
a few Gyr among the Galactic globulars and is predicated on the
assumption that age is the second parameter. In the present review, we will
examine the present state of the observational data relevant to these
predictions. First however, we summarize the 
techniques used to estimate the globular cluster ages, which are crucial
to testing the predictions.

\section{Age Determination Methods}

Our ability to estimate the ages of star clusters is founded on 
one basic principal.
Given a chemical composition, the theory of stellar evolution makes 
definite predictions about the rate of evolution of the luminosity and 
radius of a stellar model.  

Thus in principle, with perfect theoretical isochrones, and with the 
knowledge of the cluster chemical composition parameters, one 
could derive unambiguously its distance modulus and its age.  This 
is true provided one could observe either the full CMD (including 
the HB) of a star cluster, and place it in the theoretical ($M_{bol}:
log T_{eff}$) plane, or derive its luminosity function based on a complete 
sample down to magnitudes well below the main sequence turnoff. 

Practice differs from this ideal situation.  On the theoretical side, 
although we can calculate the evolution of $L$ with time, there remains 
uncertainties in the run of $R$ vs. time and in the conversion from 
$T_{eff}$ to color (Demarque et al. \cite{demea1988}; 
Bergbusch \& VandenBerg \cite{beva1992}).  
This affects not only the 
morphology of the theoretical CMD (isochrone), but also, in a much 
lesser way, the predicted luminosity function in a given spectral 
band.  In addition, the theoretical HB luminosity may be suspect
(Da Costa \& Armandroff \cite{dca1990}), 
requiring an empirical distance calibration for \mvrr.

On the observational side, there are errors in the determination of the 
CMD (in the photometry, particularly at the faint end), and in the 
cluster chemical composition, primarily \feh\  and $\rm [O/Fe]$.  
Interstellar reddening can be another source of error which matters 
in some dating methods.  In addition, when using luminosity functions, 
there are additional problems with completeness which can be very 
serious at the faint end (Bolte \cite{bol1994}; 
Sandquist et al. \cite{sanea1996}).  Finally, 
the choice of \mvrr\  is a major source of age uncertainty 
(Renzini \cite{ren1991}; Chaboyer et al. \cite{chdesa1996a}).  

In order to get around these problems, a number of methods for 
globular cluster age determinations have been devised over the 
years, which made best use of the information available at the time.  
It is important to note that there is nothing immutable about any of
these approaches.  As additional knowledge becomes available 
(either improved stellar models, or higher quality observations), 
new and better methods should be used (e.g increased use of 
luminosity functions), and approaches previously abandoned should
be reconsidered as they may become practical again.

\subsection{Main Sequence Fitting}

Main sequence fitting was the preferred method some thirty years ago.
This technique is used to derive the distance to a globular cluster,
and its age may be determined from the absolute magnitude of the
turnoff ($\rm M_V(TO)$).
Much work was done both theoretically and empirically to 
understand the dependence of the ZAMS position on chemical 
composition.  One of the important issues early-on was the uncertainty 
in the helium abundance of halo stars (Demarque \cite{dem1960}).  In this 
case, the observed unevolved main sequence in the CMD is 
superimposed on a fiducial ZAMS appropriate to the chemical 
composition of the star cluster.  In principle, the most 
straighforward way to establish the position of this fiducial ZAMS is 
to use field stars with reliable trigonometric parallaxes.  Before the 
advent of the HIPPARCOS database, this approach had 
been utilized sparingly because too few precise 
parallaxes for metal poor field dwarfs have been available 
(Carney \cite{car1980}; see also the review by Sandage \cite{san1986}).  
One could also adopt the 
position of the better established metal rich ZAMS (the Hyades main 
sequence), and apply corrections to it for metallicity differences 
(Sandage \& Eggen \cite{saeg1959}; Eggen \& Sandage \cite{egsa1962}).  
This method, 
originally based on corrections derived from the measured ultra-
violet excess $\delta$$(U-B)$ (a metallicity index), only provides an 
approximate index of the cluster metallicity and associated ZAMS 
calibration (Wildey et al. \cite{wildea1962}).  
To determine the ZAMS position, 
additional evolutionary corrections to the position of the empirical 
main sequence must be applied, which are themselves uncertain 
since they require an estimate of the ages of individual field 
subdwarfs (Eggen \cite{egg1973}).  A final problem with the main sequence 
fitting method is that it is difficult to separate reddening corrections 
due to interstellar absorption from the ultra-violet excess, both of 
which must be known separately.

For all these reasons, main sequence fitting has been generally
abondoned in recent years.  But the availability of high quality
parallaxes from the HIPPARCOS satellite, supplemented by some HST
parallaxes has revived interest in the method (Chaboyer
\ea\ \cite{chabea1997}; Gratton \ea\ \cite{gratea1997}; Pont
\ea\ \cite{pontea1997}; Reid \cite{reid1997}), since it is in principle
the most straightforward and fundamental approach.  However, even
relatively small errors in the reddening ($\pm 0.02$ mag) lead to
rather large errors in the distance modulus derived from main sequence
fitting ($\pm 0.10$ mag).  Thus, main sequence fitting is only suitable
for clusters with well determined (generally small) reddenings.  In
addition, deep main sequence photometry is required, which restricts
the use of main sequence fitting to well studied clusters.  Finally,
even with the release of the Hipparcos database, the number of
metal-poor single stars with well determined parallaxes
($\sigma_\pi/\pi < 0.10$) is rather small (less than 15).  For the
above reasons, the recent papers utilizing main sequence fitting have
concentrated on the absolute ages of a few globular clusters, and have
been unable to address the question of the relative ages of Galactic
globular clusters as a whole.

Two other globular cluster distance indicators of great potential 
importance should be mentioned.  One is the position of the white 
dwarf cooling sequence; some white dwarfs have now been 
observed with HST in the globular clusters M4 (Richer et al. \cite{ricea1995}) 
and NGC 6752 (Renzini et al. \cite{renea1996}).  The other would be the 
discovery of eclipsing binary systems near the main sequence in 
globular clusters (Paczynski \cite{pac1997}).   Both of these techniques allow
one to determine a precise distance to a cluster.  From this, the
absolute magnitude of the turnoff may be determined, and used to
estimate the age of a cluster.

\subsection{The \dv\  Method}

The \dv\  technique, where $\dv =V(MSTO) - V(HB)$
measured at the main 
sequence turnoff color, is illustrated in Fig. 1.  The main 
advantage of the \dv\ technique is that it is insensitive to 
interstellar reddening.  The quantity \dv\ can be directly measured 
from the CMD, provided the horizontal branch is sufficiently well 
defined at the color of the main sequence turnoff.  In its original 
form (Sandage \cite{san1982}) it relied on an empirical or semi-empirical 
determination of the absolute magnitude \mvrr\ of the RR Lyrae 
variables.  In recent years, the \dv\ method has been considered 
the most precise and consistent technique for dating purposes, and various 
forms of the method have been developed (Iben \& Renzini \cite{ibre1983}; 
Buonanno et al. \cite{bucofu1989}; Sarajedini \& King \cite{saki1989}; 
Chaboyer et al. \cite{chdesa1996a}).  In theory, it is also independent of 
distance since both the turnoff and horizontal branch luminosities can 
now be derived theoretically, given
the chemical composition (Sweigart \cite{swei1987},\cite{swei1994}; 
Yi et al. \cite{yilede1993}).  This is 
possible because evolutionary calculations to the helium flash and on to the
HB have been carried out for the case of no or moderate mass loss 
(Mengel \& Sweigart \cite{mesw1981}; Cole et al. \cite{codede1985}), 
applicable to the 
HB stars that evolve from red to blue. The origin of the extreme mass
loss, which is responsible for the HB stars that evolve from the
blue end of the HB (sbB and sdO), is still a matter of controversy
and several explanations have been proposed 
(Mengel et al. \cite{menogr1976}; Bailyn \& Pinsonneault \cite{bapi1995}; 
D'Cruz et al. \cite{dcea1996}). 

In practice, the \dv\  technique suffers from the difficulty in 
locating precisely the turnoff luminosity  (defined here as the bluest 
point) because the CMD sequence is nearly vertical at this point and 
difficult to measure precisely.  This difficulty may be overcome by
using a point on the subgiant branch (SGB), as discussed by 
Chaboyer \ea\ \cite{chabea1996b}.
This SGB point is +0.05 mag redder than the turnoff and represents a
location where the CMD morphology is more horizontal than the turnoff
making the magnitude measurement much more precise.
Another difficulty with the \dv\ technique concerns the 
horizontal branch luminosity [usually one adopts \mvrr]. As 
mentioned above, the HB luminosity can in principle be calculated 
theoretically since the helium core mass at the helium core flash is 
known, and such calculations are becoming increasingly reliable.  
But in recent years, the determination of \mvrr, particularly its 
dependence on \feh, has been the subject of a lively debate.  It is 
convenient to write:
\begin{eqnarray}
\mvrr\ = \alpha\,\feh  +  \beta                                  
\end{eqnarray}
where the slope $\alpha$ has important ramifications for the relative 
ages of globular clusters in the Galactic halo, and the constant $\beta$ 
in addition affects their absolute ages (Zinn \cite{zinn1985}; 
Sarajedini \& King \cite{saki1989}).  
Sandage (\cite{san1981a}, \cite{san1981b}) first derived the 
coefficients in eq.(1) 
empirically, and estimated a ``steep" slope $\alpha$ of 0.35.  From a 
theoretical calibration based on synthetic HB population models, Lee 
et al. (\cite{ledezi1990}, \cite{ledezi1994}, hereafter LDZ) 
derived a ``shallower" $\alpha$ in the 
range 0.17-0.19.  A shallow slope in good agreement with LDZ, has also been 
found using the semi-empirical Baade-Wesselink method $\alpha = 0.16\pm 0.03$
(Jones \ea\ \cite{jonea1992}) and $\alpha = 0.21\pm 0.05$ 
(Skillen \ea\ \cite{skilea1993}).
These shallow slopes have been questioned by Feast (\cite{feas1997}), who 
argued that the usual practice of treating \feh\ as the dependent ($x$)
variable was subject to bias. Feast (\cite{feas1997}) performed the inverse
regression (\mvrr\ as the dependent variable) and found $\alpha =
0.37$.  However, in performing
his regression, Feast (\cite{feas1997}) did not include the observed errors. 
We have repeated the analysis of Feast (\cite{feas1997}), 
using the same data set
but including the quoted errors in the regression.  Using a variety of
regressions (including the BCES estimator of Akritas \& Bershady
\cite{akbe1996}; FITEXY routine from Press \ea\ \cite{presea1992}; 
and the routine from
Ripley \& Thompson \cite{rith1987}), we have found the regressions (both 
forward and inverse), do not support the high value of $\alpha$
claimed by Feast (\cite{feas1997}).  Using the same 28 data points as Feast
(\cite{feas1997}), our regressions (both inverse and direct) yielded 
$\alpha = 0.22$.  The error in the slope varied from $\pm 0.04$ (BCES)
to $\pm 0.06$ (FITEXY).

The shallow slope is also supported by HST observations of globular
cluster HBs in M31, where $\alpha = 0.13\pm 0.07$ 
(Fusi Pecci et al. \cite{fusiea1996}).
Using the relation between Fourier decomposition and luminosity for
RRab stars in GCs Kov\'{a}cs \& Jurcsik (\cite{koju1996}) determined that
$\alpha$ is less than 0.20. It would appear that the evidence favors
somewhat shallow slopes ($\alpha \le 0.25$).  It is important to note 
at this point that $\alpha$ is not a universal 
constant; theory predicts that stars of the same metallicity evolve 
through the RR Lyrae instability strip at different luminosities 
depending on whether they originate on the red or blue side of the 
instability strip.  Thus $M_v(RR)$ depends on HB morphology, and 
RR Lyrae variables are more luminous in clusters with blue HB 
morphology types than in red HB morphology type, for a given 
metallicity (Lee \cite{lee1991}).  Synthetic HB population models show that 
the sensitivity of $M_v(RR)$ on HB morphology is most pronounced 
at very low metallicity.  This explains why globular cluster samples 
which contain only very blue HB types at the lowest metallicities, 
can yield a large value of $\alpha$. In addition, the value of $\beta$
is also quite uncertain, and this affects sensitively the
determination of globular cluster absolute ages 
(Iben \& Renzini \cite{ibre1983};
Chaboyer et al.  \cite{chdesa1996a},\cite{chabea1996b}).  
As a result, a semi-empirical version of
the \dv\ technique has been used in most recent discussions of
globular cluster ages, in which the turnoff luminosity is calibrated
on theoretical isochrones, and \mvrr\ is empirically derived (Chaboyer
et al. \cite{chdesa1996a}).

\subsection{The \dbv\ Method}

The quantity \dbv\ is defined as the color difference between 
the turnoff and the base of the giant branch (Sarajedini \& Demarque 
\cite{sade1990}; VandenBerg et al. \cite{vabost1990}, VBS; 
Sarajedini \cite{sar1991}).  It is illustrated in 
Fig. 1.  Because it requires a well defined turnoff/subgiant branch and a 
fine grid of isochrones for the chemical composition of the cluster, it 
has only been introduced recently in globular cluster dating.  Since 
the position of the giant branch in the CMD is to first order only a 
function of chemical composition, \dbv\ decreases with 
increasing age.  In practice, the method is calibrated using the \dv\
method, but in principle, a set of theoretical isochrones could 
provide an absolute age calibration.

The \dbv\ method is excellent for measuring the relative ages of 
clusters  with similar
metallicities.  For absolute ages, its calibration suffers from 
uncertainties in the radii of stellar models which are sensitive to the 
depth of the convection zone and to the efficiency of helium and heavy 
element diffusion in the envelope during evolution (Deliyannis et al. 
\cite{dedeka1990}; Demarque et al. \cite{dedesa1991}; 
Chaboyer et al. \cite{chsade1992}). Furthermore,
unlike the \dv\ method whose calibration with age is largely
insensitive to the particular isochrones used, ages
determined with \dbv\ can vary widely depending on which isochrones
are used for the calibration.

A slight variation on the ``classical" \dbv\ method has recently been
introduced by Saviane et al. (\cite{saropi1997}). They measure the $V-I$ color
difference between the main sequence turnoff and the red giant branch
(RGB) at a level 2.2 magnitudes brighter than the MSTO. It is unclear
what advantages this has over \dbv\. However, because it combines
the concept of the \dbv\ method with knowledge about the magnitude of
the MSTO, the method of Saviane et al. (\cite{saropi1997}) may be subject to
additional observational and theoretical errors.

\subsection{The Use of Luminosity Functions}

In principle, luminosity functions could be used to determine the 
cluster age uniquely, without any external knowledge of the cluster 
distance or interstellar reddening, provided the chemical composition 
is known.  It is in practice difficult to achieve; accurate photometry 
is needed down to faint magnitudes well below the turnoff in order to 
determine precisely the change in slope of the luminosity function 
due to the turnoff, and for the same reason completeness of the data 
is also essential down to faint magnitudes.  Because of these 
difficulties, luminosity functions have until recently not been used 
for dating globular clusters.  Some recent studies of globular cluster 
giant branch luminosity functions (Jimenez \& Padoan 1996) can 
provide useful self consistency checks of evolutionary rates along 
the giant branch, but not an absolute calibration of cluster ages.  
However, with the availability of high precision photometry, 
luminosity functions near the turnoff and down the main sequence 
will become increasingly useful for globular cluster dating.  This is 
particularly so because the turnoff slope change in luminosity 
functions is less sensitive to details of the theoretical shape of the 
isochrone in the CMD, which are affected by convection zone depth 
and element diffusion rates.

\subsection{How Good Are Stellar Models Near The Main 
Sequence Turnoff?}

Our best information comes from the Sun. The extraordinary 
precision of the Global Oscillations Network Group (GONG) data indicate 
that the standard solar model 
predicts the correct run of the sound speed in the Sun to one part in 
one thousand, provided the effects of helium and heavy element 
diffusion are included in the models.  Perhaps more importantly in 
our context, the small spacings of the solar p-modes now permit for 
the first time the determination of the solar age from seismology (4.5 
Gyr) (Guenther \& Demarque \cite{gude1997}); and it agrees to within 0.1 Gyr 
with the radioactive age of the Sun (Guenther \cite{guen1989}; 
Bahcall et al. \cite{bapiwa1995}).  This means that in the case of 
the Sun and its early evolution near the main sequence, the standard 
theory predicts the solar age to within two per cent, and it will soon 
be possible to do much better.  One might mention also that in the 
case of two nearby field disk population binary systems, Procyon 
($\alpha$ Canis Minoris) and $\alpha$ Centauri, conventional stellar 
models are consistent with the available observations on masses and 
chemical compositions to better than ten per cent.  It is reasonable to 
expect that as for the Sun, seismology will provide very stringent 
tests of the internal structure, distances, and ages of these two 
binary systems 
and of a few nearby dwarfs and subgiants in the field (e.g. the 
recent observations of $\eta$ Bootes by Kjeldsen et al. \cite{kjelea1995}).

Thus recent results for the Sun are very encouraging, but at this date 
stellar seismology is still in the future.  Thus until we can rely on 
good quality seismic data for a number of nearby dwarfs and 
subgiants, the ages of old open star clusters must be determined by 
conventional theoretical isochrone fitting, and can be determined at 
best to within ten per cent.  Observationally, the errors are due to the 
difficulty in defining the main sequence precisely because of small 
numbers and cluster membership problems, and errors in distances 
and interstellar reddening.  On the theoretical side, uncertainties are 
primarily due to convection, convective core overshoot and the 
depth of the convective envelope.

The problem of the depth of the convection zone, coupled with the 
uncertainties of the treatment of convection in model atmospheres 
for halo stars, and the associated uncertainties in the transformation 
from theoretical temperatures and gravities to observed colors are 
significant caveats to our reliance on the detailed shapes of 
theoretical isochrones near the main sequence turnoff.   Helium and 
heavy element diffusion are also known to play a role in affecting
the shape of the turnoff (Noerdlinger \& Arigo \cite{noar1980}; 
Proffitt \& Michaud \cite{prmi1991}; Chaboyer et al. \cite{chsade1992}).   
Much basic work remains 
to be done in the area of numerical simulations of radiative 
hydrodynamics as applied to stellar envelopes and atmospheres in 
low metallicity stars (Fuhrmann et al. \cite{fuaxge1993}; 
van't Veer-Menneret \& 
Megessier \cite{vame1996}).  In all these areas, progress will be aided by 
recent advances in helio- and astero-seismology (e.g. Demarque et 
al. \cite{deguki1997a}).  These uncertainties argue strongly  in favor 
of using the 
\dv\ method, which is less dependent on the least known aspects 
of stellar evolution.

Halo stars in the field and in globular clusters are generally 
more distant than 
disk stars, and we do not know as much about their individual 
properties.  There are however a few factors which facilitate the 
precise dating of globular clusters.  Observationally, because of 
large numbers and no cluster membership problems, the main 
sequence can be remarkably well delineated (see e.g. NGC 6752, 
Rubenstein \& Bailyn \cite{ruba1997}), and with some effort, reliable 
luminosity functions can be derived (Sandquist et al. {sanea1996}).   On the 
theoretical side, low metallicity means better determined opacities 
and equation of state than in the Sun and Sun-like stars, and 
therefore probably more reliable isochrones.  Despite these 
advantages, we must remember that empirical data on the masses 
and luminosities of individual metal-poor dwarfs are sorely needed.  
The Hipparcos mission has provided high quality parallaxes for 
$\sim 15$ individual subdwarfs in the field, and parallaxes from HST may
increase this number in the future.  Perhaps good fortune will also 
lead to the discovery of eclipsing binaries near the turnoff of 
globular clusters. 

The different age determination methods make use of
different sets of information from GC photometry, and by and large
make use of only part of the information contained in the data.  The
result is that different methods can yield different age calibrations.
The discrepancies are indicators of the uncertainties in the stellar
models and the observations, and much can be learned from them about 
the details of stellar
physics.  For example, uncertainties in stellar radii will affect the
\dv, \dbv\ and luminosity function calibrations in different ways.  As
the data become more complete and more accurate to faint magnitudes,
they will provide increasingly stringent tests of stellar evolution
theory.

These then are the various techniques used to estimate the ages of
globular clusters. We now proceed with our discussion of what the
globular cluster ages tell us about the formation of the Galactic halo.
First, the nature of the second parameter effect is examined. We then
go on to discuss the age range among the Galactic globulars and the
existence of an age-metallicity relation.

\section{The Concept of Age as the Global Second Parameter}

The SZ scenario for the formation of the Galactic halo is based upon the
premise that cluster age is the global second parameter, but what exactly
does `global second parameter' mean?
The work of SZ and, more recently, Lee et al. (\cite{ledezi1994}) has shown 
that
the second parameter effect varies with Galactocentric distance. Globular
clusters inside 8 kpc of the Galactic center are minimally affected by
the second parameter, if at all, while those outside 8 kpc are significantly
affected. That is to say, whatever this second parameter actually is, it 
varies on a global scale. Some have argued that there may not
be a single second parameter affecting the HB morphology (SVB). In this
scenario, the global variation in HB-type is caused by different parameters 
acting in {\it precisely} the proper amount to yield the observed variation of
HB-type. This situation seems rather contrived and is therefore unlikely
to be correct. The large scale (global) variation of the second parameter
effect argues in favor of a global second parameter. 

Let us assume now that there is a global second parameter, which along with
metallicity, influences the HB morphologies of globular clusters. We must
also assert that it is highly unlikely that there exist two second parameters,
which affect the HB morphology to {\it precisely equal} degrees. What is more
likely is that there is the first parameter (metallicity), the second
parameter, the third parameter, and so on. As a result, it makes little
sense to speak of `a' second parameter, which implies that more than one
parameter of equal significance is at work.

\subsection{The Case of NGC 288 and NGC 362}

Ever since Sandage \& Wildey (\cite{sawi1967}) proposed that the 
second parameter is
helium abundance and SZ made convincing arguments in favor of age as the
second parameter, a number of investigators have endeavored to reveal the
nature of the global second parameter. The primary question that is asked
is: ``Can a difference in age account for the difference in HB-types of clusters
at a given metal abundance?" As noted above, M3 and M13 were
recognized early-on as being a `second parameter pair' in which two clusters
have similar metallicities but different HB-types. However, these were not
the first such clusters to be studied in detail. The first
pair of clusters to receive in-depth scrutiny are NGC 288 and NGC 362.
These have $[Fe/H] \approx -1.3$, but NGC 288 has a completely blue HB
whereas NGC 362 has an HB that is predominantly red. 

Bolte (\cite{bol1989})
presented the first detailed comparison of the main sequence turnoff
(MSTO) regions of NGC 288/362. He constructed precise B--V color-magnitude 
diagrams (CMDs) for these clusters down to 3 magnitudes below the MSTO. 
Then, by shifting them by the relative reddenings and distance moduli,
both of which are small, Bolte (\cite{bol1989}) showed that the MSTO of NGC 362
is brighter than that of NGC 288. This leads directly to an age difference
of $\approx$3 Gyr with NGC 288 being older than NGC 362. This finding, along
with the work of King et al. (\cite{kidegr1988}) and 
Dickens et al. (\cite{dicea1991}), revealed 
direct unequivocal evidence that the HB of NGC 288 is bluer than that of 
NGC 362 because the former is older than the latter. 

A number of other authors have reached the same conclusion (e.g. Demarque
et al. \cite{demea1989}; Sarajedini
\& Demarque \cite{sade1990}; VandenBerg et al. \cite{vabost1990}). 
Most notable among these is the work of Green \& Norris (\cite{grno1990}, 
hereafter GN90). They followed an
approach similar to that of Bolte (\cite{bol1989}) except that they 
chose to work in
the B--R CMD arguing that changes in the morphology of the CMD near the
MSTO are more pronounced in B-R as compared with B--V. The conclusion
reached by GN90 is identical to that of Bolte (\cite{bol1989}),
namely that NGC 288 is older than NGC 362 by 3 Gyr. In contrast, 
VandenBerg \& Durrell (\cite{vadu1990}) used the magnitude of the first ascent
red giant branch tip to estimate the relative distance moduli of NGC 288
and NGC 362 to allow
a comparison of the MSTO's of these clusters. From such a comparison,
which included the adoption of relative reddenings, they concluded that 
there is no detectable age difference between these
clusters. However, in considering the significance of this result, one
must weigh the following points. GN90 note that the brightest first
ascent giant star in NGC 288 should have a similar color to an analogous
star in NGC 362 if the metallicities of the clusters are indeed similar.
However, GN90 argue that the colors of the stars are in fact significantly
different; this and other evidence presented by GN90 indicates that 
VandenBerg \& Durrell have chosen a star that is several tenths of a 
magnitude too faint to be at the giant branch tip. Da Costa \& 
Armandroff (\cite{dca1990}) have reached a similar conclusion after considering 
the bolometric magnitudes of the brightest stars on the RGB of NGC 288.
Therefore, it seems that if one modifies the analysis of VandenBerg \& Durrell
to account for the improperly chosen giant branch tip magnitude of NGC 288, one
again reaches the conclusion that NGC 288 is older than NGC 362 by
$\approx$3 Gyr.

Until recently, it seemed that a general consensus was building regarding
the relative age of NGC 288/NGC 362. The review presented by SVB includes
a new analysis of the entire question using NGC 1851, a cluster with both
a red and blue HB, as a `bridge' between NGC 288 and NGC 362. In the
spirit of VandenBerg \& Durrell (\cite{vadu1990}), they once 
again concluded that an age difference between these clusters is not proven. 
They begin by assuming that all three clusters (NGC 288,
NGC 362, NGC 1851) have the same age. This permits the registration of
their main sequence/subgiant branch (SGB) fiducial sequences using the nearly 
horizontal portion of the SGB as shown in Fig. 2a. This procedure 
yields magnitude and color offsets required to bring the three sequences into 
coincidence at the SGB. If any age differences exist between the clusters,
then using these magnitude and color offsets on the brighter portion of
the CMD should cause their HBs to be misaligned. Fig. 3 shows a comparison
of the NGC 1851 fiducial sequence from Walker (\cite{walk1992}) 
with the photometric data of NGC 362 (Harris \cite{harr1982}) and data 
for NGC 288 from Bolte (\cite{bol1992}) and Bergbusch (\cite{berg1993}). 
The latter has been offset to the photometric scale of 
Bolte (\cite{bol1992}). The comparisons in Fig. 3 indicate that while the 
fiducial sequence of NGC 1851 
lines up adequately with the photometric data for NGC 362, it
does not faithfully reproduce the position and shape of the NGC 288
RGB and HB. In particular, the red end of the blue HB is fainter in
NGC 1851 as compared with NGC 288, and the entire blue HB of NGC 1851
is systematically brighter/bluer than that of NGC 288. 
This is contrary to the conclusions drawn by SVB from a similar
comparison. The difference in results is due to the fact that SVB 
included NGC 1851 RR Lyraes in their definition of the position of this
cluster's blue HB thus extending the red end of the blue HB into the
RR Lyrae instability strip. In fact, assuming $E(B-V)=0.02$
(Da Costa \& Armandroff \cite{dca1990}), the intrinsic color of the blue edge of
the instability strip in NGC 1851 is inferred to be $(B-V)_0=0.23$ from
the blue HB envelope drawn by SVB in their Fig. 7. From the work of
Sandage (\cite{san1990}), this is approximately 0.05 mag too red since he places
the blue edge of the strip at $(B-V)_0 = 0.18$. Furthermore, it turns out
to be extremely difficult to find magnitude and color offsets that allow the
entire fiducial sequence of NGC 1851 to fit the photometric data of NGC 288.
We return to this point below.
As for the apparent alignment of the NGC 362 and 1851 HBs in Fig. 3, what 
is not shown in Fig. 6 of SVB is the poor correspondence of the NGC 362 
and NGC 1851 unevolved main sequences between $V\approx20.2$ and
$V\approx21.5$ now seen in Fig. 2a. If we adjust the offset for NGC 362 to optimize the fit 
in this magnitude range, we arrive at the comparison shown in Fig. 2b. 
This clearly shows that when the main sequences of NGC 288 and 362 are
aligned, the latter is younger than the former. In addition, with this
new offset for NGC 362, its red HB will be brighter than that of NGC 1851
as it should be given that the luminosity of the red HB is dependent
on age as well as metallicity (see Sarajedini et al. \cite{salele1995},
hereafter SLL). 
It appears therefore that 
the work of many previous investigators presenting evidence in favor of 
an age difference between NGC 288 and NGC 362 continues to stand on a robust
foundation. 

As mentioned above, it is difficult to satisfactorily align the entire
fiducial sequence of NGC 1851 from the RGB to the HB to the main sequence
with the photometry for NGC 288. It is unclear whether this is due to
an age difference, a chemical abundance difference, a combination of these
two effects, or something completely unknown. What is clear is that 
NGC 1851 is an extremely unusual cluster in that its HB morphology is
strongly bimodal. In addition, Catelan (1996, private communication) has 
found that the
RR Lyraes in NGC 1851 possess unusual period shifts.

\subsection{The Case of M3 and M13}

The difference in HB-types between M3 and M13 is not as extreme as the
difference between NGC 288 and NGC 362. As a result, if age is the global
second parameter, then we would expect the age difference between M3 and
M13 to be somewhat less than $\approx$3 Gyr. One of the first modern 
attempts to answer this question was provided by VandenBerg et al. (1990).
Although their photometry was of sufficient precision, they could not
unequivocally address the age difference question because there were only
a handful of stars on the subgiant and red giant branches in the CMDs
making the measurement of $\Delta$$(B-V)$ uncertain. Their quoted
age differences, though frought with uncertainties, did indicate that
M3 may be younger than M13 by 1 to 2 Gyr. This is consistent in sign and
magnitude with what we would expect based on the comparison between
NGC 288 and 362. 

Catelan \& de Freitas Pacheco (\cite{cafr1995}) applied the $\Delta$$V$ 
method to M3 and
M13. They combined published values of $\Delta$$V$ for each cluster with
synthetic HBs designed to estimate the RR Lyrae luminosity. This latter
approach is especially important in the case of M13 which has a very blue
HB morphology and few RR Lyrae variables. Based on this analysis, they
are able to set an upper limit of $\approx$3 Gyr on the age difference
between M3 and M13. Such a limit is not a useful constraint when one expects
an age difference less than $\approx$3 Gyr. Another factor to consider
when evaluating the
work of Catelan \& de Freitas Pacheco (\cite{cafr1995}) is that they utilized
cluster photometry with a precision that is inadequate to answer the question
they posed. For example, in the case of M3, they relied on the $\Delta$$V$
value estimated by Buonanno et al. (\cite{bucofu1989}) 
which actually dates back to 
the photographic photometry of Buonanno et al. (\cite{buonea1986}) 
and some apparently
unpublished data. Catelan \& de Freitas Pacheco examined two different
possibilities for the $\Delta$$V$ value of M13 both of which are 
problematic. In one case, photographic
photometry of only 3 RR Lyraes from Pike \& Meston (\cite{pime1977}) is 
used to estimate
V(HB), and in the other case, the CCD photometry of 
Guarnieri et al. (\cite{gubrfu1993}),
which exhibits significant photometric scatter, is utilized to yield the 
turnoff magnitude. Thus, the results of their analysis do not serve to
shed light on the question of whether there is an age difference between
M3 and M13 because the quality of the observational inputs is inadequate.

\subsection{The Case of the Galactic Globular Clusters as a Whole}

In addition to examining the relative ages of second parameter cluster 
pairs, it is also important to investigate the Galactic globular cluster
system as a whole. In this way it is possible to see if the age
explanation is adequate to explain the global behavior of the second
parameter effect. A number of investigators have considered the variation
of cluster age with HB morphology over a narrow range in metallicity. 
Most of this 
work has used the $\Delta$$V$ age diagnostic. For example, 
Sarajedini \& King (\cite{saki1989}) and Chaboyer et al. (\cite{chsade1992}) 
conclude that there is a statistically significant correlation between
age and HB type with older clusters having bluer HB types. 
Carney et al. (\cite{castjo1992a}) have performed a similar 
analysis using different
observational and theoretical inputs and reached the same
conclusion though at a lower level of significance. 

Zinn (\cite{zinn1993}) has adopted a different approach to 
this question. He used the
$[Fe/H]$:HB type diagram to divide the Galactic
globulars into ``old halo'' and ``younger halo'' subsamples based on their
position in this diagram (see his Fig. 1). He found a correlation
between cluster age (as derived from both the $\Delta$$V$ and 
$\Delta$$(B-V)$ methods) and HB type with the younger clusters
having redder HBs. Chaboyer et al. (\cite{chdesa1996a}) have taken 
Zinn's (\cite{zinn1993}) 
analysis further by determining new and improved $\Delta$$V$ ages for 43
Galactic globulars and computing the mean ages of the old halo and
younger halo samples. These age designations were made based on the
expectation that redder HB clusters are younger, and indeed the $\Delta$$V$
ages of Chaboyer et al. (\cite{chdesa1996a}) indicate that the 
younger halo clusters
are some 2 to 4 Gyr younger than the old halo clusters with statistical 
significances in the 4$\sigma$ to 8$\sigma$ range. Even when the globular
clusters belonging to the Sagittarius dwarf galaxy and those
with very young ages are all removed, the age difference between
the younger halo and the old halo falls in the range 1.5 to 2.5 Gyr
with a mean significance level of 3.4$\sigma$. 

We can take these analyses still further by avoiding the step of converting
measured $\Delta$$V$s to ages and simply examining the behavior of
$\Delta$$V$ with HB type. The top panel of Fig. 4 shows the canonical
$[Fe/H]$:HB type diagram for the Galactic globular clusters using data 
from Lee et al. (\cite{ledezi1994}, hereafter LDZ94). The solid
lines are the results of synthetic HB calculations (LDZ94) and indicate 
the expected locations of clusters for different relative ages
based on the assumption that age is the sole global second parameter.
In the lower panel of Fig. 4, we investigate the behavior of $\Delta$$V$ 
with HB type for two metallicity ranges: $-1.6 \leq [Fe/H] \leq -1.4$
(filled circles) and $[Fe/H] \leq -2.0$ (open circles) utilizing the
observational database of Chaboyer et al. (\cite{chdesa1996a}). 
Weighted least squares fits to the filled circles yield highly 
significant relations (probabilities $>$ 99\%) even if we randomly 
remove single points from the fits. This is also true when fits are 
performed to the open circles. The solid line in the
lower panel is the expected behavior of  $\Delta$$V$ as inferred from
the synthetic HB models (top panel) for clusters with $[Fe/H] = -1.5$ while the
dashed line is the expected behavior for clusters with $[Fe/H] = -2.2$.
These have been shifted vertically to match the mean locations of the
observed $\Delta$$V$ values. These lines reproduce the
general trends quite well providing further
evidence that age is the strongest candidate to be the global second
parameter.

\subsection{Second Parameter Effect Among Red HB Clusters}

The principal diagnostic diagram used to gauge the second parameter effect
has traditionally been the aforementioned $[Fe/H]:(B-R)/(B+V+R)$ diagram. 
However, as several authors have noted, this diagram becomes insensitive
to the ``second parameter" when the number of blue HB stars becomes
zero (i.e. $B=0$). That is to say, one cannot explore the existence of a
second parameter among red HB clusters using the $[Fe/H]:(B-R)/(B+V+R)$ plane.
However, one way in which we {\it can} explore this issue is by plotting 
$[Fe/H]$ as a function of the dereddened mean color of the red HB stars
($\langle(B-V)_{o,RHB}\rangle$),
for clusters with $(B-R)/(B+V+R) < -0.8$.
Figure 5a shows such a diagram which includes the observational
data analyzed in SLL (filled circles), 
consisting of Galactic globular clusters as well as 
the SMC cluster Lindsay 1 and the LMC cluster ESO121-SC03. In addition,
the globular cluster Pyxis (Sarajedini \& Geisler \cite{sage1996})
and the SMC clusters Kron 3 and NGC 121 (Mighell et al. \cite{misafr1997}, 
all open
circles) are also plotted in Fig. 5a. The conversion from $(B-R)_o$
to $(B-V)_o$ for Pyxis has been performed using the empirical transformation
devised by Sarajedini \& Geisler (\cite{sage1996}). The adopted reddenings are taken from the compilation of Harris (\cite{harr1996}), and the papers 
by Rich et al. (\cite{ridamo1984})
for Kron 3, Stryker et al. (\cite{stdamo1985}) for NGC 121, and Sarajedini \&
Geisler (\cite{sage1996}) for Pyxis. Also plotted in Fig. 5 are the 
mean intrinsic
colors of Zero Age Horizontal Branches (ZAHBs) for scaled-solar
abundances and masses of 0.90$M_\odot$ (left dashed line) and
0.66$M_\odot$ (right dashed line) from the work of Dorman (\cite{dorm1992}).
These lines are meant to indicate the general behavior of ZAHB color with
$[Fe/H]$ as well as with mean mass.

The dominant influence of the first parameter, metallicity, is evident 
in Fig. 5a with more metal-rich clusters having generally redder intrinsic
HB colors. In addition, there appears to be a nonzero width in the 
distribution of colors at a given metallicity. This is the classic
signature of the presence of a second parameter effect. One weakness
of this diagram is the need to adopt a reddening for each cluster. This
adds additional uncertainty to the color axis of Fig. 5a. To circumvent
this problem, Fig. 5b shows $[Fe/H]$ as a function of
the difference in color between the RGB and red HB. The horizontal error
bars are estimated by adding, in quadrature, the standard error of the mean HB 
color and the standard error of the RGB color. This procedure yields
a mean error in $d_{B-V}$ of $\approx$0.01 mag. In accord with the
conclusions drawn from Fig. 5a, Fig. 5b also indicates
that there is a significant spread in $d_{B-V}$ at a given metallicity,
especially in the range $-1.5\lea[Fe/H]\lea-1.0$. 

The most straightforward
explanation for this nonzero spread in $d_{B-V}$ is the presence of an
age range among these red HB clusters. Thus, at a given metal abundance,
clusters with smaller values of $d_{B-V}$ (or larger values of 
$\langle(B-V)_{o,RHB}\rangle$) are younger. In fact, SLL
demonstrated that age differences estimated from $d_{B-V}$ are fully
commensurate with relative ages inferred from the position of the 
MSTO for clusters with metalliciites in the range where their
theoretical models were applicable. In the present paper, we can perform
a similar analysis looking specifically at the relative values of
$\langle(B-V)_{o,RHB}\rangle$ between clusters of similar metallicity.
We have chosen the cluster NGC 362 with $[Fe/H] \approx -1.3$ 
(Zinn \& West \cite{ziwe1984}) as the
comparison cluster. Fiducial sequences are available for NGC 362 in 
$B-R$ (Green \& Norris \cite{grno1990}) and 
$B-V$ (Harris \cite{harr1982}; VandenBerg et al. \cite{vabost1990}).
Figures 6 and 7 show CMD comparisons of six red HB clusters with the
fiducial sequence of NGC 362. The photometric data are taken from the
following sources: Kron 3 (Rich et al. \cite{ridamo1984} [$B-R$]; 
Mighell et al. \cite{misafr1997} [$B-V$]), 
NGC 121 (Stryker et al. \cite{stdamo1985} [$B-R$]; 
Mighell et al. \cite{misafr1997} [$B-V$]), 
Pyxis (Sarajedini \& Geisler \cite{sage1996}), 
Palomar 4 (Christian \& Heasley \cite{chhe1986}), 
Lindsay 1 (Olszewski et al. \cite{olaasc1987}), 
and ESO121-SC03 (Mateo et al. \cite{mahosc1986}). 
Each cluster has been registered to the magnitude
of the NGC 362 red HB and the color of the RGB at the magnitude of the HB.
This is not strictly correct since SLL showed that the luminosity of
the red HB is dependent on metallicity and age, but for the purposes
of the present comparison, making this correction will not alter the
results.
It is immediately obvious from Figs. 6 and 7 that there is a correlation
between relative HB color and relative age (as judged from the MSTO 
magnitude) with respect to NGC 362. In particular, the clusters with 
brighter MSTO mags also have redder HB colors in comparison with NGC 362.

\subsection{Exceptions To the Rule?}

\subsubsection{Arp 2}

The globular cluster Arp 2, which is believed to be a member of the
Sagittarius dwarf galaxy, has $(B-R)/(B+V+R) = +0.86$ 
(Sarajedini \& Layden \cite{sala1997}, hereafter SL97), a 
predominantly blue HB morphology, even though
Buonanno et al. (\cite{buonea1995}, hereafter BCFRF) claim 
that its age is some 3 Gyr younger than other clusters at its metallicity 
($[Fe/H] = -1.80 \pm 0.10$, SL97). 
As a result, the properties of Arp 2 clearly contradict
the explanation of age as the sole global second parameter. However, 
we note that the measured $(B-R)/(B+V+R)$ value of Arp 2 is
an upper limit since investigators have been forced to assume $R=0$ because
of the severe field contamination present in this part of the CMD.
To gauge the importance of this effect, we have counted stars in the
red HB portion of the SL97 Arp 2 CMD; we find
$R = 9 \pm 3.0$, assuming Poisson statistics. In an identical region of
their `off-cluster' CMD, we count $3 \pm 1.7$ stars yielding a net value of
$R = 6 \pm 3.4$. Keeping in mind that $B = 24 \pm 4.9$ and $V = 4 \pm 2.0$
(SL97), we calculate 
$(B-R)/(B+V+R) = 0.53 \pm 0.17$. This value is not as anomalous as
the value obtained under the assumption that $R=0$, and we have shown that,
indeed, $R$ is not likely to be identically zero.

The age of Arp 2 is another matter that requires in-depth scrutiny.
BCFRF quote a value of $\Delta$$V = 3.29 \pm 0.10$
for the magnitude difference between the MSTO and the HB and
$\Delta$$(B-V) = 0.248 \pm 0.005$ for the color difference
between the MSTO and the lower RGB. When compared with the $\Delta$$V$
values of CDS and the $\Delta$$(B-V)$ values given by VBS, both of 
these quantities indicate
that Arp 2 is a relatively young cluster. However, we arrive at a
different conclusion when we reanalyze their photometry. Figure 8
shows the BCFRF Arp 2 photometry compared with the
fiducial sequence of M68 (Walker \cite{walk1994}) 
registered by matching the magnitude of
the HB and the RGB color at the level of the HB. This is done because
both clusters have very similar metallicities 
(M68 has $[Fe/H] = -2.09 \pm 0.11$; Zinn \& West \cite{ziwe1984}).
First, we note that $V(HB) = 18.20 \pm 0.04$ for Arp 2, which is composed 
of $18.13 \pm 0.04$ (RR Lyrae mag from SL97) plus $0.074 \pm 0.002$ 
(offset from SL97 zeropoint to that of BCFRF). We adopt $V(HB) = 15.64$
for M68 (Walker \cite{walk1994}). What is immediately obvious from Fig. 8 is
that the apparent $\Delta$$V$ of Arp 2 does not
differ drastically from that of M68. The vertical
dashed lines in Fig. 8 show the measured
$\Delta$$(B-V)$ value (as defined by VBS) quoted by BCFRF. 
Inspection of Fig. 8 is enough to cast doubt on the finding
that Arp 2 is significantly younger than other clusters at its
metallicity. A similar conclusion results from a comparison of Arp 2 with
the fiducial sequences of M92 and NGC 6397.

It is possible to use the new age estimator described by Chaboyer et al.
(\cite{chabea1996b}) to calculate a more precise age for Arp 2 
using the point brighter
and +0.05 mag redder than the MSTO [$M_V(BTO)$]. Utilizing the
procedure outlined by Chaboyer et al. (\cite{chabea1996b}), we find 
$V(BTO) = 21.25 \pm 0.02$, which, when coupled with the HB mag given
above, yields $\Delta$$V(BTO) = 3.05 \pm 0.04$. Assuming
$[Fe/H] = -1.80 \pm 0.10$ (SL97), the models of
Chaboyer et al. (\cite{chabea1996b}) give an age of $14.4 \pm 0.7$ Gyr 
for Arp 2, which can be compared with $15.2 \pm 0.4$ Gyr for their 
17 metal-poor 
globular clusters. These ages agree to within the errors, but
there is the possibility that Arp 2 could be slightly younger than
other clusters at its metallicity (perhaps $\approx$1 Gyr), but certainly
not $\approx$3 Gyr younger. When considered along with the downward-revised
$(B-R)/(B+V+R)$ value, the age of Arp 2 is no longer a glaring
exception to the hypothesis that age is the sole global second parameter.

\subsubsection{IC 4499}

Ferraro et al. (\cite{ferrea1995}, hereafter FFFCB) present a 
CCD CMD of IC 4499 in the
$BV$ passbands. They quote a value of $\Delta$$V = 3.25 \pm 0.12$. From 
this and their measured $\Delta$$(B-V)$ value,
they conclude that IC 4499 is 3-4 Gyr younger than other globulars at its
metallicity, which is $[Fe/H] = -1.75 \pm 0.20$. This large age difference
does not seem to match the HB morphology of IC 4499 measured to be
$(B-R)/(B+V+R) = 0.08$ by Sarajedini (\cite{sara1993}). 
For comparison, M3 also has
$(B-R)/(B+V+R) = 0.08$ and $[Fe/H] = -1.66 \pm 0.06$ 
(Zinn \& West \cite{ziwe1984}). 
As a result, there is an apparent
conflict between the HB type of IC 4499, which matches that of M3, and its
age which appears to be much younger than that of M3.

Using the IC 4499 photometry of Ferraro et al. (\cite{ferrea1995}) divided 
into radial
bins to optimize the definition of the CMD sequences, we measure
$V(BTO) = 20.54 \pm 0.01$. Together with the HB magnitude of
$V(HB) = 17.65 \pm 0.04$, we calculate $\Delta$$V(BTO) = 2.89 \pm 0.04$
for IC 4499. This compares favorably with $\Delta$$V(BTO) = 2.92$
for M3 estimated from the fiducial sequence of 
Buonanno et al. (\cite{buonea1994}). 
Thus, IC 4499 does not seem to be significantly younger than M3, 
thereby refuting the existence of any contradiction between the HB
and age of IC 4499.

We should point out however that IC 4499 possesses a rather
bizarre and intriguing HB. First of all, it has the highest specific frequency 
of RR Lyrae
variables among Galactic globular clusters. Walker \& Nemec (\cite{wane1996})
found that $66 \pm 5$\% of the stars on the HB are RR Lyraes of one
form or another. Sarajedini (\cite{sara1993}) found that the color extent of the
BHB is significantly smaller than those of M3 and NGC 3201. Based on
this, he speculated that the mean mass of the
BHB stars in IC 4499 is at least $\approx$0.02$M_{\odot}$ higher than 
those in more normal clusters such as M3 and NGC 3201.

\subsubsection{Bimodal Horizontal Branch Clusters}

There are three Galactic globular clusters whose HBs appear to be 
the result of having combined a purely blue HB, like that in NGC 288,
with a purely red HB, like that of NGC 362. These bimodal HB clusters are
NGC 1851 (Walker \cite{walk1992}) and NGC 2808 (Byun \& Lee \cite{byle1991};
Sosin et al. \cite{sosea1997}). In addition,
Borissova et al. (\cite{borisea1997}) have claimed that NGC 6229 also has
a bimodal HB. The reason for these peculiar HBs is as yet unclear.
Lee et al. (\cite{ledezi1988}) claim that they are the natural 
result of evolution
away from the ZAHB, while others have evoked more exotic phenomena.

More recently, Rich et al. (\cite{richea1997}) have
obtained CMDs in the cores of several clusters using the Hubble Space
Telescope and WFPC2. In the case of three
clusters thought to have purely red HB morphologies (NGC 362, NGC 6441,
and NGC 6388), they have discovered a small fraction of blue HB stars.
The number of blue HB stars is between 10\% and 15\% of the total number
of red HB stars. In contrast, the ground-based CMDs of NGC 6388 
(Silbermann et al. \cite{silbea1994}) and NGC 362 (Harris \cite{harr1982}), 
which mostly
probe the outer regions of these clusters, exhibit very weak blue HBs,
if any, as compared with the HST CMDs. Therefore, this variation in 
the HB morphology near the cluster center is likely to be a result of
some sort of stellar dynamical effect; it may be connected to increased
mass loss due to stellar interactions or it may be related to the
subdwarf-B phenomenon studied by Bailyn et al. (\cite{bailea1992}). 
In either case, 
their apparent affect on HB morphology, which has 
in the past introduced some confusion in discussions of the 
systematics of HB morphology, are now understood to reflect local 
conditions rather than a stage in the evolution of isolated single 
stars, and are probably not relevant to the global properties of the 
globular cluster system.

\subsubsection{Other Candidates For the Global Second Parameter}

In addition to age, there are a number of other globular cluster 
star properties that could be
considered to be candidates for the global second parameter. These are
individually discussed and effectively discounted by the arguments presented
in LDZ94. For example, there is absolutely no possiblity that a change
in the CNO abundance could explain the behavior seen in Figs. 6 and 7.
Although increasing the CNO abundance makes the HB morphology redder,
it also makes the MSTO redder and fainter, which is the opposite of what
is observed. Another example is a decrease in the helium abundance, which
makes the HB redder as well, but also causes the MSTO to become redder
and fainter, which is also not what is observed. The reader is referred to 
LDZ94 and Demarque et al. (\cite{demea1989}) for more examples.

Buonanno et al. (\cite{buonea1997}) have recently published the 
latest installment in a series of papers (Buonanno \cite{buon1993}; 
Fusi Pecci et al. \cite{fusiea1993}; 
Buonanno \& Iannicola \cite{buia1995}; 
Fusi Pecci et al. \cite{fusiea1996}; see also 
van den Bergh \& Morris \cite{vamo1993}) which examine the role
of stellar density on the morphology of the HB. Their first conclusion
is that clusters having similar metallicity but different central stellar
densities tend to exhibit different blue HB morphologies. In
particular, Buonanno et al. (\cite{buonea1997}) claim that clusters with higher 
central densities are more likely to populate the bluest extremes
of the HB. Their second conclusion is illustrated in Fig. 9. The top
panel (9a) shows the variation of $B2/B+V+R$ with the peak dereddened
color of the HB [$(B-V)_{peak}$]. The HB-type index $B2/B+V+R$ is composed
of the number of stars blueward of $(B-V)_o = -0.02$ (i.e. the blue HB tail), 
denoted by B2, divided by the sum of the number of stars blueward of the 
instability strip (B), redward of the instability strip (R), and the number 
of RR Lyrae variables (V). As would be expected, Fig. 9a shows that
when the HB peaks in the blue ($(B-V)_{peak}$$\lea$0.3), $B2/B+V+R$ tracks
the peak HB color quite well. However, redward of this color, $B2/B+V+R$ 
becomes insensitive to the peak HB color because the quantity
B2 goes to zero. In order to model the general variation of 
$B2/B+V+R$ with $(B-V)_{peak}$, Buonanno et al. (\cite{buonea1997}) performed a
least squares fit to the filled circles in Fig. 9a; this yielded the
solid line shown in the same figure, which likely represents the effect
of the first parameter (metallicity) on the HB morphology. Then, they 
plotted the residuals
from this fit versus the log of the central density of each
cluster (Log $\rho_o$) as shown in Fig. 9b. The apparent correlation
between the $B2/B+V+R$ residuals and the central density seems to indicate
that at a given $(B-V)_{peak}$, the strength of the blue HB is controlled 
by the stellar density in the cluster. This is in-line with the findings 
of Djorgovski et al. (1997) discussed above. The dashed line in Fig. 9a
shows the `inverse' linear fit (i.e. independent and dependent variables 
are interchanged) to the data points with $(B-V)_{peak}$$\lea$0.3. This
inverse fit appears to represent the general behavior of $B2/B+V+R$ with 
$(B-V)_{peak}$ much better than the direct fit. Fig. 9c illustrates 
the residuals from this inverse fit as a function of Log $\rho_o$. In
this case, there is no compelling reason to believe that these residuals
are correlated with the cluster central density. As a result, this casts
doubt on the role that central density plays in influencing the HB
morphology.

In this section, we have presented arguments in favor of the idea that
age is the global second parameter. However, this does not mean that 
some of the other second parameter candidates play no role whatsoever in
influencing the HB morphologies of a minority of clusters. As with all 
general trends, there are a few exceptions to the rule.   

\section{The Age Range}

Detailed studies of individual GCs have demonstrated rather convincingly
that a few GCs are substantially younger (or older) than other
clusters with a similar metallicity (e.g.\ Rup 106, Buonanno \ea 
\cite{buonea1993}; Pal 12, Stetson \ea\ \cite{stetea1989}; 
NGC 362, see previous section).
However, a few anomalously young (or old) GCs do not answer the
question of whether the bulk of the outer halo formed as the result
of a rapid collapse (ELS), or over an extended period of time (SZ).
This requires a study of the age range among the bulk of
the Galactic GCs.  The \dbv\ technique cannot be reliably applied to
clusters of differing metallicity, so one must examine GC ages
obtained via \dv\ in order to explore the existence of an age range
among the bulk of Galactic GCs.  We performed such a study of the \dv\
ages of 43 GCs in CDS, and concluded that a total ($\pm 2\sigma$) age
range of 5 Gyr exists among the bulk of the Galactic GCs.  

This work was severally criticized by SVB, who suggested that CDS
underestimated the Gaussian 1-$\sigma$ error bars associated with the
\dv\ measurements.    Although observers typically quote errors in their
derived \dv\ (or V(TO), V(HB) ) values, they never state if these are
to be interpreted as Gaussian 1-$\sigma$ error bars.  Indeed, it is
very difficult for an observer who makes a single measurement of
\dv\ to estimate the true Gaussian 1-$\sigma$ error associated with
the measurement.  Walker (1992) provides an example of the difficulty
in estimating the error; he originally estimates $\dv = 3.42\pm 0.05$
(page 582), ``but given the difficulty in measuring the turnoff to an
accuracy better than $\pm 0.10$ mag'' gives $\dv = 3.42\pm 0.10$ mag
in his abstract.  CDS used the larger error bar in their
data table.  

CDS attempted to directly estimate the Gaussian 1-$\sigma$ error bar
associated with measurements of \dv\ by examining independent
measurements of \dv\ made by various authors.  An analysis of these
repeated observations led CDS to conclude that the spread among these
observations was much smaller than would be expected if the observers
were quoting Gaussian 1-$\sigma$ error bars.  A reasonable estimate
of the Gaussian error could be obtained by multiplying the quoted
errors by 0.61.  However, SVB pointed out that a few of the `independent'
\dv\ measurements actually could be traced back to similar sources.
For this reason, we have re-checked all of the original references,
removed those that were in error and repeated our analysis.  This
re-analysis is presented in Appendix A, where it is determined that 
multiplying the quoted errors by 0.61 does indeed lead to a reasonable
estimate of the true 1-$\sigma$ error  associated with  the \dv\
measurements tabulated by CDS. 

An independent estimate of the Gaussian error associated with
measuring \dv\ can be made by considering all of the old, metal-poor
clusters.  One would expect that these clusters should all have the
same \dv.  Hence, the standard deviation about the mean should be
similar to the average Gaussian error in the individual measurements.
An old, metal-poor GC sample may be selected independent of
\dv\ by selecting GCs which have  $\feh \le -1.6$; have not been 
shown to be young using \dbv\  and are not suspected of
being young based on HB morphology (i.e. belonging to the Old Halo
group of Zinn \cite{zinn1993}).  Applying these selection criteria to the \dv\
measurements in Table 2 of CDS along with the additional requirement
that the original authors have included an estimated error in their
\dv\ measurement results in a list of 12 metal-poor, old globular
clusters with \dv\ measurements: NGC 2298, 5024, 5897, 6205, 6254,
6341, 6397, 6535, 6809, 7078, 7099, and 7492.  The average \dv\ error
quoted by the observers is $\pm 0.158$ mag.  However, the standard
deviation of the points about the mean is $\sigma = 0.080$ mag,
calculated using the standard formulae $\sigma^2 \equiv \Sigma(x_i -
\overline{x})^2/(N-1)$. Alternatively, one may use the small number
statistics of Keeping (1962), whose formulae also yield $\sigma =
0.080$ for the 12 metal-poor clusters.  This standard deviation is
remarkably similar to the average 1-$\sigma$ \dv\ error bar used by
CDS ($\pm 0.083$ mag), suggesting that CDS used the appropriate
1-$\sigma$ error in their analysis.
 
For these old, metal-poor clusters, the standard deviation about the
mean is significantly smaller (factor of 2) than the average error
quoted by the observers, suggesting that observers are not quoting
Gaussian 1-$\sigma$ error bars.  This may be quantified by comparing
the observed distribution of measured \dv\ values to an expected
distribution.  The expected distribution is constructed by randomly
generating 1000 \dv\ values for each input \dv\ assuming the mean
value given by the observer, and using a Gaussian distribution with
the $\sigma$ of the Gaussian taken to be the observer's error bar.  In
this test, the expected distribution has $\sigma = 0.17$ and the
$F$-test (Press \ea\ \cite{presea1992}) finds that there is less than a 1\%
probability that the observed distribution has the same standard
deviation as the expected distribution.  When the observed error bars
are multiplied by 0.61 (as advocated by CDS), then the expected
distribution has $\sigma = 0.12$ (still larger than the observed
$\sigma$).  There is a 10\% probability that the observed and expected
distributions have the same standard deviation. Thus, the measured
\dv\ values for old, metal-poor GCs strongly support the conclusion of
CDS, that a reasonable estimate of the Gaussian 1-$\sigma$ error
associated with the measurement of
\dv\ can be estimated by assuming the errors quoted by observers
correspond to $\sim 1.64\,\sigma$ error bars.  This essentially implies
that the quoted error ranges correspond to 90\% confidence levels, as
opposed to the 68\% confidence level for a Gaussian 1-$\sigma$ error.

As an additional justification of their criticism of CDS, SVB give
some anecdotal remarks about the observational difficulties associated
with measuring \dv.  We fully agree with this, as mentioned in \S 2.2.
Even with good photometry, the turnoff region can be nearly vertical
for $\sim 0.2$ mag.  However, this does not imply that the Gaussian
1-$\sigma$ error in the measured turnoff must be of order 0.1 mag.
The average 1-$\sigma$ \dv\ error bar used by CDS was $\pm 0.083$ mag.
Assuming that the error in \dv\ is due in equal parts to the error in
determining the HB and TO levels, this implies that 68\% of the time,
observers will be able to determine the magnitude of the TO within
$\pm 0.06$ mag.

Finally, SVB based part of their criticism on a quote from 
Grundahl (\cite{grun1996}, PhD thesis), who compared the \dv\ ages of 
CDS with the \dbv\  ages of Richer \ea\  (\cite{ricea1996}).  
Richer \ea\ (\cite{ricea1996}) determined relative GC ages
within a given metallicity group using \dbv, and derived the absolute
age of the different metallicity groups using a reference cluster in
each group whose age was derived from isochrone fits to the observed
color magnitude diagram.  According to SVB, Grundahl concluded that
there was a poor match between the Richer \ea\  (\cite{ricea1996}) ages and the
CDS ages, and the likely cause were the errors in the adopted \dv\
values from the literature.  However, it is not at all clear how
Grundahl performed such a comparison, as Richer \ea\  (\cite{ricea1996}) 
{\em did
not give any errors in their derived ages}.  The Richer
\ea\  (\cite{ricea1996}) ages are based on fitting isochrones to observed color
magnitude diagrams.  As discussed in Section 3 above, the
colors predicted by stellar models and isochrones are very uncertain,
due to the difficulties in treating convection, and our incomplete
knowledge of stellar atmospheres.  Thus, ages derived by fitting
isochrones to GC color magnitude diagrams will have very large
uncertainties attached to them.  It is not surprising then, that
Grundahl (\cite{grun1996}) did not find a good match between the Richer \ea\ 
(\cite{ricea1996}) ages and the CDS ages.

Salaris et al. (\cite{sawe1997}, hereafter SW) have recently completed a study
which relies upon relative \dbv\ ages within 4 metallicity groups and
uses \dv\ ages to define the absolute age of a reference cluster in
each metallicity group.  The study of SW has an additional advantage
over the work of Richer \ea\  (\cite{ricea1996}) of quoting errors in their
derived ages.  Comparing the 24 CDS and SW ages in common\footnote{In
performing this comparison, we have used the CDS ages derived
assuming $\mvrr = 0.20\feh + 0.82$.  This is the slope preferred by
CDS.  The adopted zero-point leads to a rather young mean age (though
still older than the mean age found by SW). CDS found that the derived
age range did not depend critically on the adopted \mvrr-\feh\
relation.}  reveals a linear Pearson R correlation coefficient of
0.707, with a probability of correlation of 99.99\%. This clearly
reveals an excellent correlation between the \dv\ ages of CDS and the
mainly \dbv\ ages of SW.  SW derived rather young ages, and the mean
ages of the two samples are different.  However, when examining the
question of an age range, it is the relative ages that are important.
A linear fit between the SW and CDS ages (which takes into account the
errors in both variables, Press \ea\  \cite{presea1992}) 
yields a reduced $\chi^2$
of 0.85, which, if anything suggests that the error bars are
overestimates.  If the two very young clusters are removed from the
fit (Ter 7 and Pal 12), a correlation still exists, and the reduced
$\chi^2$ is still less than 1 (0.90), indicating a good fit.  If one
uses the observer's original error bars in the CDS ages (as advocated
by SVB), then the reduced $\chi^2$ is 0.57 for all ages in common.  On
average, such a small value of the reduced $\chi^2$ only occurs 5\%
of the time, suggesting  that the fit is `too good'.  This is caused by
overestimating the errors. Of course, when comparing ages for
individual GCs between CDS and SW, discrepancies may be found.
However, CDS did not concern themselves with individual ages, but the
properties of the sample as a whole.  As such, the excellent
correlation between the CDS ages and SW ages provides yet another
piece of evidence which supports the errors adopted by CDS and the
conclusion that a total age range of $\sim 5$ Gyr exists among the
bulk of the Galactic GCs.\footnote {At the time of this writing,
the absolute age scale of globular clusters is in a state of flux, but 
if absolute ages are indeed reduced by some factor, age 
differences would be reduced by approximately the same factor.}

\section{Age and Metallicity}

The question of whether an age-metallicity relation exists among the
Galactic GCs is a longstanding one.  As the relative \dbv\ ages are
only reliable over a small metallicity range, ages derived using \dv\
are usually used to examine the correlation between age and
metallicity among GCs.  It has been traditionally assumed that \dv\ is
independent of metallicity, leading to the conclusion that the
existence of an age-metallicity relation depends critically on the
slope of the \mvrr-\feh\ relation (Sandage \cite{san1982}; 
Buonanno \cite{bucofu1989}; Sarajedini
\& King \cite{saki1989}; Sandage \& Cacciari \cite{saca1990}). 
However, as shown by CDS
there are now a number of clusters whose \dv\ values are smaller than
the average (e.g.\ Rup 106, Pal 12, Ter 7).  As none of the known
`young' GCs are metal poor ($\feh \la -1.8$), CDS found that an
age-metallicity relation exists even for relatively steep \mvrr-\feh\
slopes of 0.30.  However, this correlation is driven by the small
number of young clusters, and CDS found that the a global
age-metallicity relation only exists when the \mvrr-\feh\ slope is less
than 0.26.  This study was based on individual \dv\ measurements, and
somewhat older stellar models.

The large majority of \dv\ measurements compiled by CDS are for $\feh
< -1.2$.  If we restrict our attention to these relatively metal-poor
clusters, and require that the clusters belong to the `Old Halo' group
of Zinn (\cite{zinn1993}), then an examination of the \dv\ values reveals that
they are indeed consistent with a single mean value, independent of
\feh.  The linear Pearson R correlation coefficient between \feh\ and
\dv\ is 0.166 (22 points), with a probability of correlation of
53\%.  Thus, among these old, relatively metal-poor clusters the
assumption that \dv\ is independent of \feh\ is valid.

Assuming that \dv\ is independent of \feh\ (in the range $-2.4 \le \feh
\le -1.2$) allows for a simple analytical analysis of the
age-metallicity relation. The basic assumptions are equation (1):
$\mvrr = \alpha\feh + \beta$, and
\begin{equation}
{\rm M_V(TO)} = c\,t + d\,\feh + e
\label{eqmvto}
\end{equation}
where $\alpha,\beta,c,d$ and $e$ are constants.  The constants in equation 
(\ref{eqmvto}) can be derived from theoretical isochrones.  Note that
CDS found that a more complicated quadratic relation was required to
accurately model the behavior of the theoretical isochrones.  This
was due to the relatively large \feh\  ($-2.5$ to $-0.5$ dex) and age
(8 - 22 Gyr)  range used by CDS.  For the restricted \feh\  and age
range (10 - 16 Gyr) considered here, the linear relation given in 
(\ref{eqmvto}) is adequate.   Combining equations (1) and 
(\ref{eqmvto}) along with the definition of \dv\ yields
\begin{equation}
t = \frac{\dv}{c} - \frac{d-\alpha}{c}\feh - \frac{e-\beta}{c}.
\label{eqage}
\end{equation}
The age-metallicity relation is simply defined by the derivative of
the above equation
\begin{equation}
\frac{dt}{d\feh} = \frac{\alpha-d}{c}
\label{eqagefeh}
\end{equation}
which assumes that \dv\ is independent of \feh.  From this, it is
clear that the existence of an age-metallicity relation depends both
on the theoretical isochrones (through $c$ and $d$) and on the slope
of the \mvrr-\feh\ relationship (the coefficient $\alpha$).  Previous work
has focused on the later point.  Here, we address both points, by
considering a variety of isochrones and \mvrr-\feh\ slopes.  

The Monte Carlo isochrones of 
Chaboyer \ea\ (\cite{chdesa1996a},\cite{chabea1996b}) provide a large
set (1000) of isochrones which have been calculated under a wide
variety of physical assumptions (different oxygen abundances, mixing
lengths, opacities, etc).  For each set of isochrones, the
coefficients $c$ and $d$ were calculated and combined with a variety
of \mvrr-\feh\ slopes to derive an age-metallicity relation.  The
results for 4 slopes are presented in Table 1.  For an
\mvrr-\feh\ slope of 0.2, Table 1 shows that the most
likely value for the age-\feh\ slope is $-1.6\,$Gyr/dex, and the 95\%
confidence limits on this bound are $-1.0$ to $-2.0$.  Thus, if the
\mvrr-\feh\ slope is 0.20, then GCs at $\feh = -1.2$ are $1-2$ Gyr
younger than $\feh =-2.2$ GCs.
\begin{table}[t]
  \begin{center}
  \begin{tabular}{cccc}
\multicolumn{4}{c}{TABLE 1}\\
\multicolumn{4}{c}{AGE-METALLICITY RELATION}\\[3pt]
\hline\hline
&&\multicolumn{2}{c}{95\% confidence limits}\\[-5pt]
\multicolumn{1}{c}{$\displaystyle \frac{d\mvrr}{d\,\feh}$}&
\multicolumn{1}{c}{$\displaystyle \frac{d\,t}{d\,\feh}$}&
\multicolumn{1}{c}{Upper}&
\multicolumn{1}{c}{Lower}\\
\multicolumn{1}{c}{(mag/dex)}&
\multicolumn{1}{c}{(Gyr/dex)}&
\multicolumn{1}{c}{(Gyr/dex)}&
\multicolumn{1}{c}{(Gyr/dex)}\\[2pt]
\hline
$0.15$ & $-2.2$ & $-2.8$ & $-1.8$\\
$0.20$ & $-1.6$ & $-2.0$ & $-1.0$\\
$0.25$ & $-0.6$ & $-1.4$ & $-0.1$\\
$0.30$ &\hspace*{1.0em}$0.0$ &\hspace*{0.1em}$+0.5$ & $-0.6$\\
\hline
\end{tabular}
\end{center}
\label{tabagefeh}
\end{table}

Alternatively, one may determine the
critical \mvrr-\feh\ slope, above which no age-metallicity relation
exists.  Based upon the typical errors in determining $c$ and $d$, a
minimum slope of $0.50\,$Gyr/dex is required to conclusively
demonstrate the existence of an age-metallicity slope.  Using the
Monte Carlo isochrones, the mean
critical value of the \mvrr-\feh\ slope is 0.26, above which no
age-metallicity relation exists (very similar to that found by CDS).
The 95\% confidence limits on this slope are 0.23 and 0.30.
Thus, even \mvrr-\feh\ slopes as shallow as 0.23 may not lead to an
age-metallicity relation, depending on which isochrones  are used.  

The exact value of the \mvrr-\feh\ slope (denoted by $\alpha$) is
still open to debate (see \S 2.2).  It would appear that the evidence
favors somewhat shallow slopes ($\alpha \le 0.25$). However, one
cannot rule out slopes around $\alpha =0.23$.  Thus, based upon the
Monte Carlo analysis presented in the previous paragraph, one cannot
conclusively state whether or not an age-metallicity relationship
exists among the old, metal-poor Galactic GCs.

Finally, there is some evidence that $\alpha$ is not the same in all 
metallicity ranges, and that even within a given metallicity range, 
$\alpha$ changes with time as the HB type evolves from red to blue.  
This last non-linearity is most pronounced for the most metal-poor 
clusters, for which $M_v(RR)$ varies by a larger amount during 
evolution (Lee \cite{lee1991}; Demarque et al. \cite{demea1997b}).  
This additional 
complexity makes the existence of an simple age-metallicity 
relationship somewhat problematic.

\section{How DID the Milky Way Form?}

Given all of the points in the above discussion, it seems to us that the
most viable formation model for the Milky Way is one that combines the
properties of the ELS and SZ scenarios as suggested by several previous
authors (e.g. Norris \cite{norr1994}). Recall that the ELS model was
based on the kinematics of nearby stars and is therefore only applicable
to the region inside of $\sim$10 kpc from the Galactic plane. In this region,
ELS maintained that the age range among the globular clusters is very small.
This assertion is actually in agreement with one of the basic tenets of the
SZ model, namely that inside of 8 kpc from the Galactic center, no second
parameter effect exists. In other words, the age range among the halo globular clusters inside 8 kpc is very small. Furthermore, inside of $\sim$8 kpc,
there is a relation between \feh\ and Galactocentric distance for the halo
globular clusters (Zinn \cite{zinn1986}). 
All of these characteristics point to a 
monotonic collapse {\it a la} ELS inside of $\sim$8 kpc of the Galactic center. 

In contrast, outside of 8 kpc, there appears to be a significant age
range among the Galactic globulars of as much as 5 Gyr. Additionally,
the second parameter effect becomes stronger as Galactocentric distance
increases indicating the need for a larger age spread to explain the wide
range of HB morphologies at a given metallicity. Furthermore, there is
no relation between \feh\ and Galactocentric distance. All of these 
observations point to a prolonged chaotic collapse highlighted by
the fragmentation of the proto-galactic gas cloud. 
In this manner, we reconcile the two competing paradigms for the formation
of the Galactic halo. 

\section{Summary}

The relative ages of the Galactic globular clusters provide an
insightful look into the earliest formation epochs of the Galaxy's
halo. In particular, studying these ages can allow us to distinguish
between a rapid collapse of the halo characterized by the scenario of
ELS or a more gradual collapse similar to the one described by SZ. 

To begin with, we need measurement techniques that will yield globular
cluster ages to the required level of precision. These methods can be
classified into four general groups: 1) Direct measures of the cluster
distance, hence age, via main sequence fitting, 2) Vertical measures
which utilize the magnitude difference between various features in the
color-magnitude diagram, 3) Horizontal measures which use the color
difference between CMD features, and 4) measures which rely on counting
stars in various stages of evolution, i.e. using the luminosity
function. 

Through the use of these age-determination methods, we have examined various
questions concerning the Galactic globular clusters. First, we investigate
the second parameter effect in detail. Weighing all of the evidence
presented in numerous previous works as well as new analyses described 
herein, we conclude that the age of a globular cluster is the
most likely candidate to be the global second parameter, which along with
metal abundance, controls the morphology of the horizonal branch.

Next, we address the question of an age range among the globular clusters
using the \dv\ age diagnostic. The answer to this question is greatly
dependent on obtaining a robust estimate for the observational errors in 
the measured \dv\ values of clusters. We present arguments
in favor of the assertion that the quoted errors in \dv\ represent 90\% confidence intervals, and that Gaussian 1-$\sigma$
errors are approximately 60\% of the quoted errors. When the measured \dv\ 
values for
43 globular clusters are combined with these updated errors, we find
a total age range of $\sim$5 Gyr among the bulk of these clusters.

Finally, we use theoretical isochrones to derive an algebraic expression 
which describes the age - metallicity relation. The slope of this equation
determines whether there is a relation between age and metallicity.
Monte Carlo simulations of theoretical isochrones indicate that 
there is a significant relation between age and
metallicity among the Galactic globular clusters if the slope of the
\mvrr-\feh\ relation is less than $\sim$0.20.

\acknowledgments

We are grateful to Doug Geisler, Caty Pilachowski,  Taft Armandroff,
Bob Zinn, and Richard Larson
for providing comments on an earlier
version of this manuscript.  Ata Sarajedini was supported by NASA grant
number HF--01077.01--94A and Brian Chaboyer was supported by NASA grant
number HF--01080.01--96A from the Space Telescope Science Institute,
which is operated by the Association of Universities for Research in
Astronomy, Inc., under NASA contract NAS5-26555.

\clearpage

\figcaption[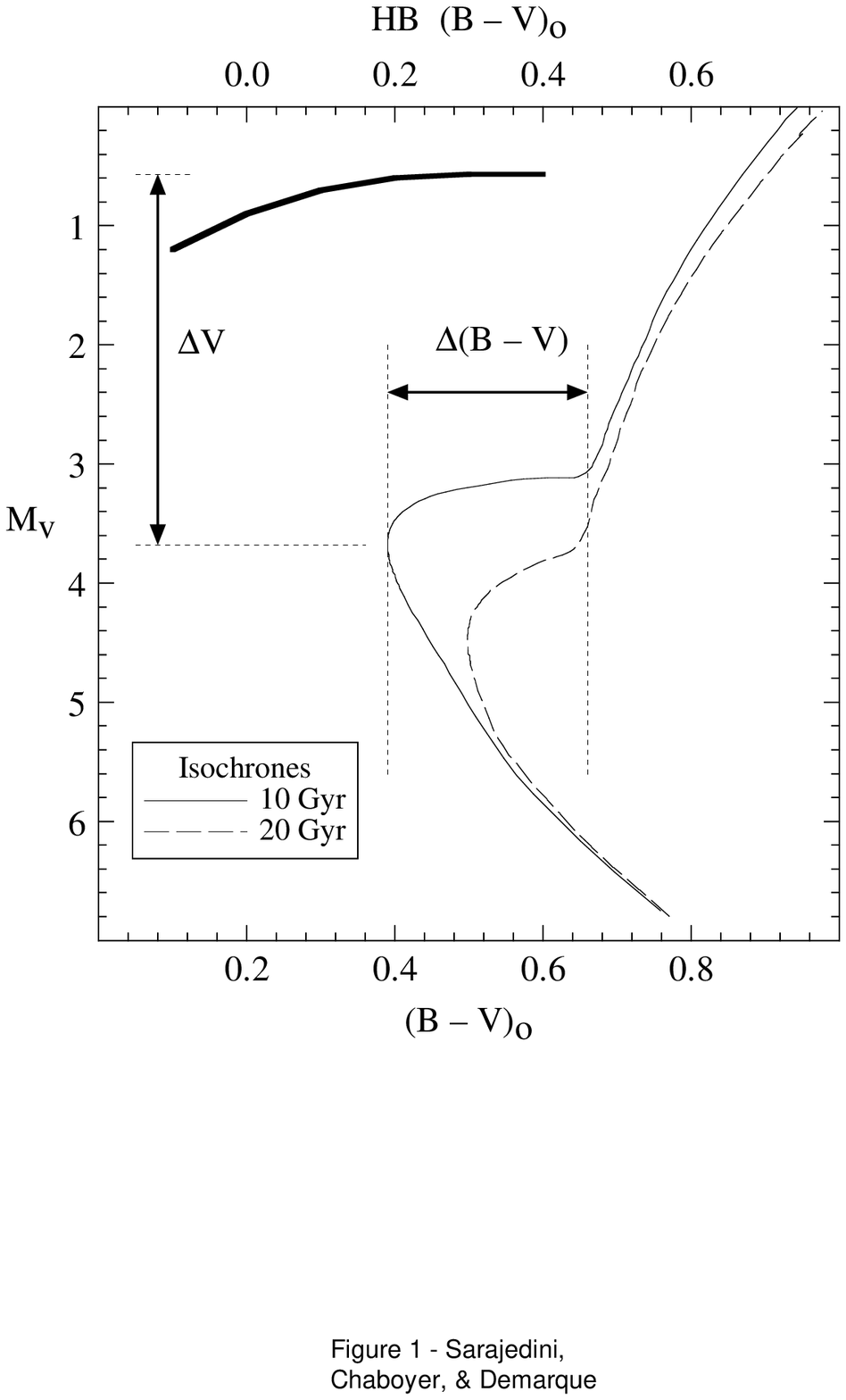]{Definitions of the \dv\ and \dbv\ age diagnostics
discussed in the text. Note that the magnitude level of the horizonal
branch is essentially constant with age. \label{fig1}}

\figcaption[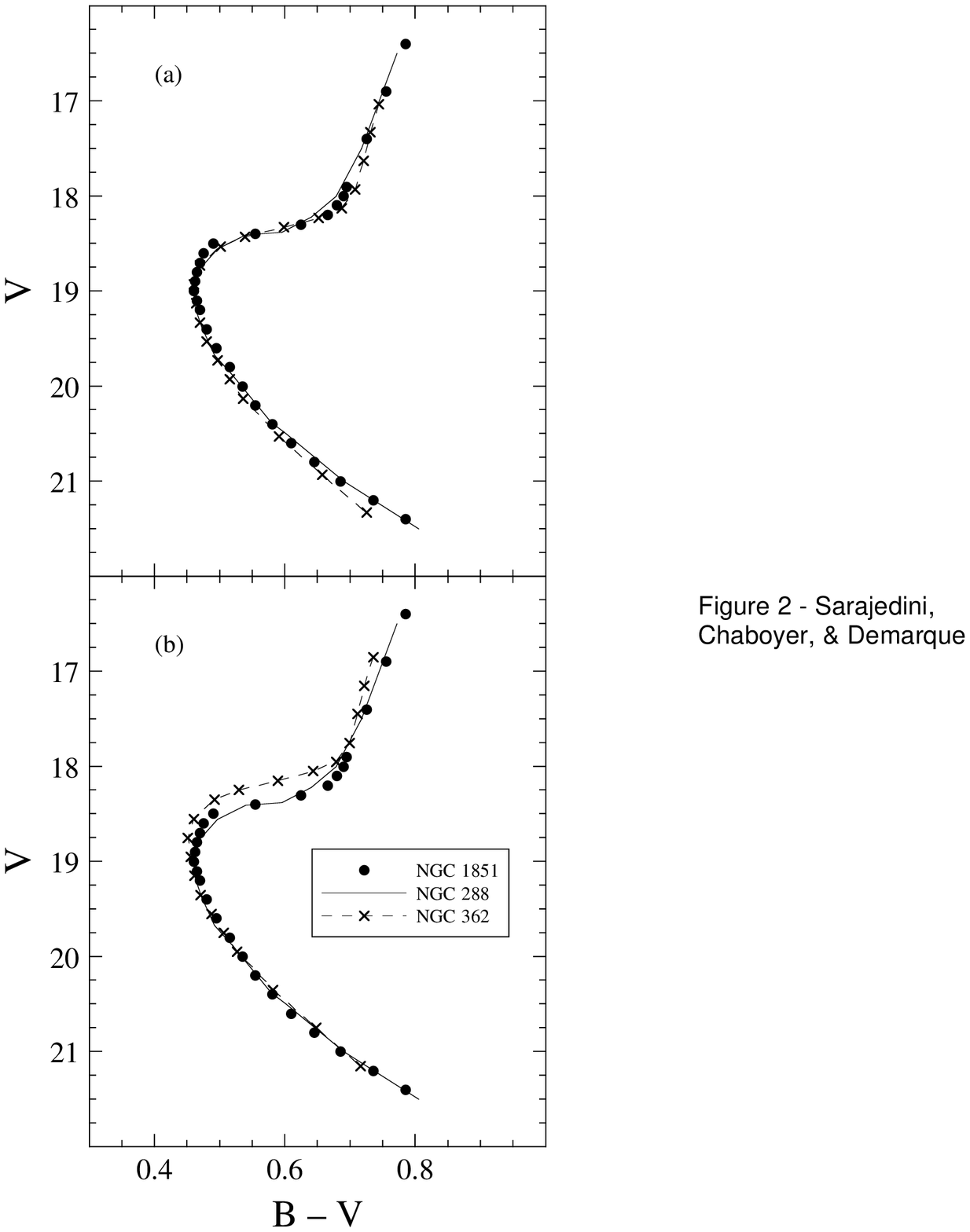]{(a) The fiducial sequences of NGC 362 (x's),
NGC 1851 (filled circles), and NGC 288 (solid line) are offset to
match the horizontal portion of the subgiant branch. (b) Same as in
(a) except that the fiducials have been shifted to match the unevolved
main sequences. \label{fig2}}

\figcaption[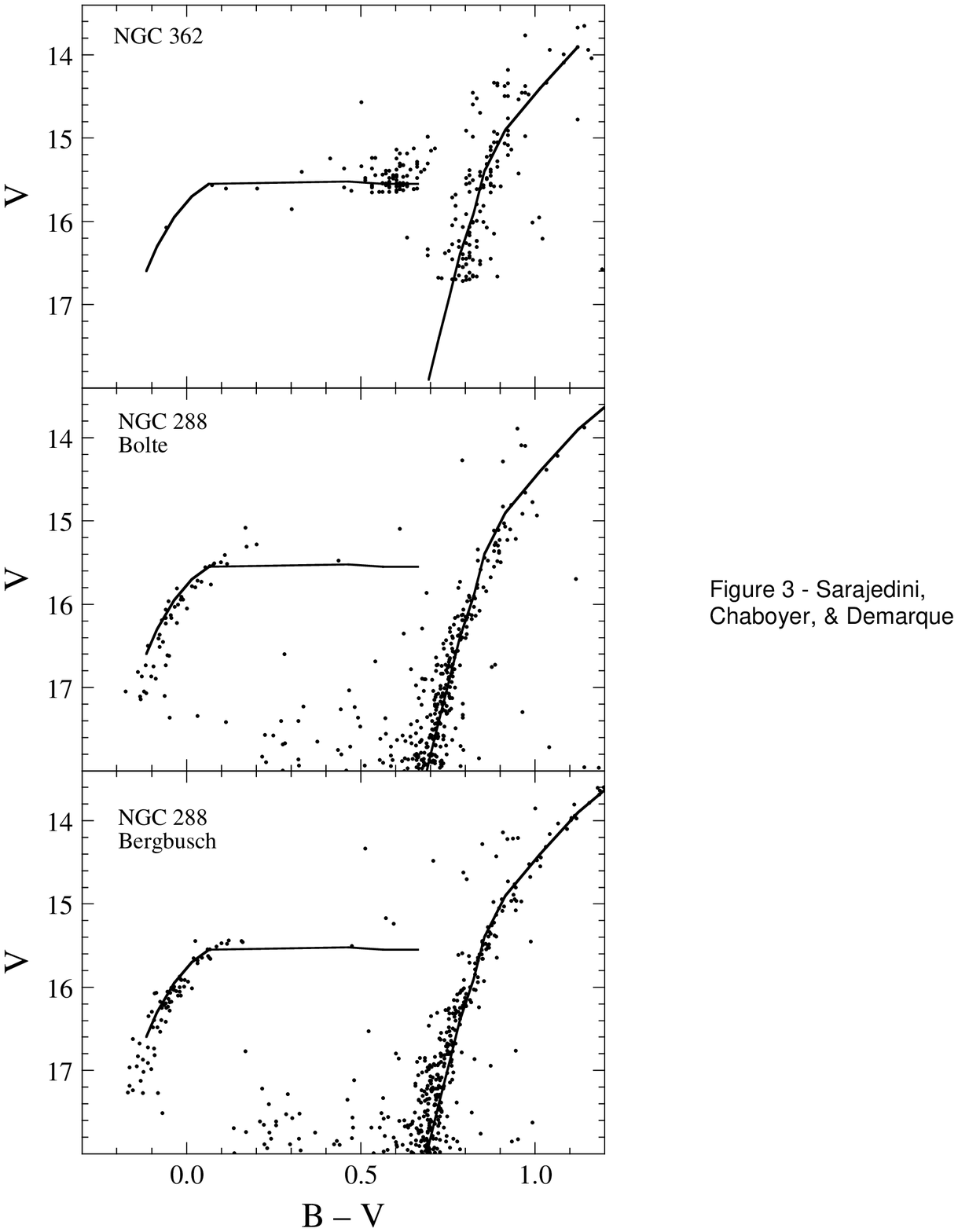]{Utilizing the magnitude and color offsets that
match the cluster subgiant branches (see Fig. 2a), the top panel 
shows a comparison of the NGC 362 photometry with the NGC 1851 fiducial
sequence. This same fiducial is compared with the NGC 288 photometry
from Bolte (1992) in the middle panel and Bergbusch (1993) in the
the lower panel. \label{fig3}}

\figcaption[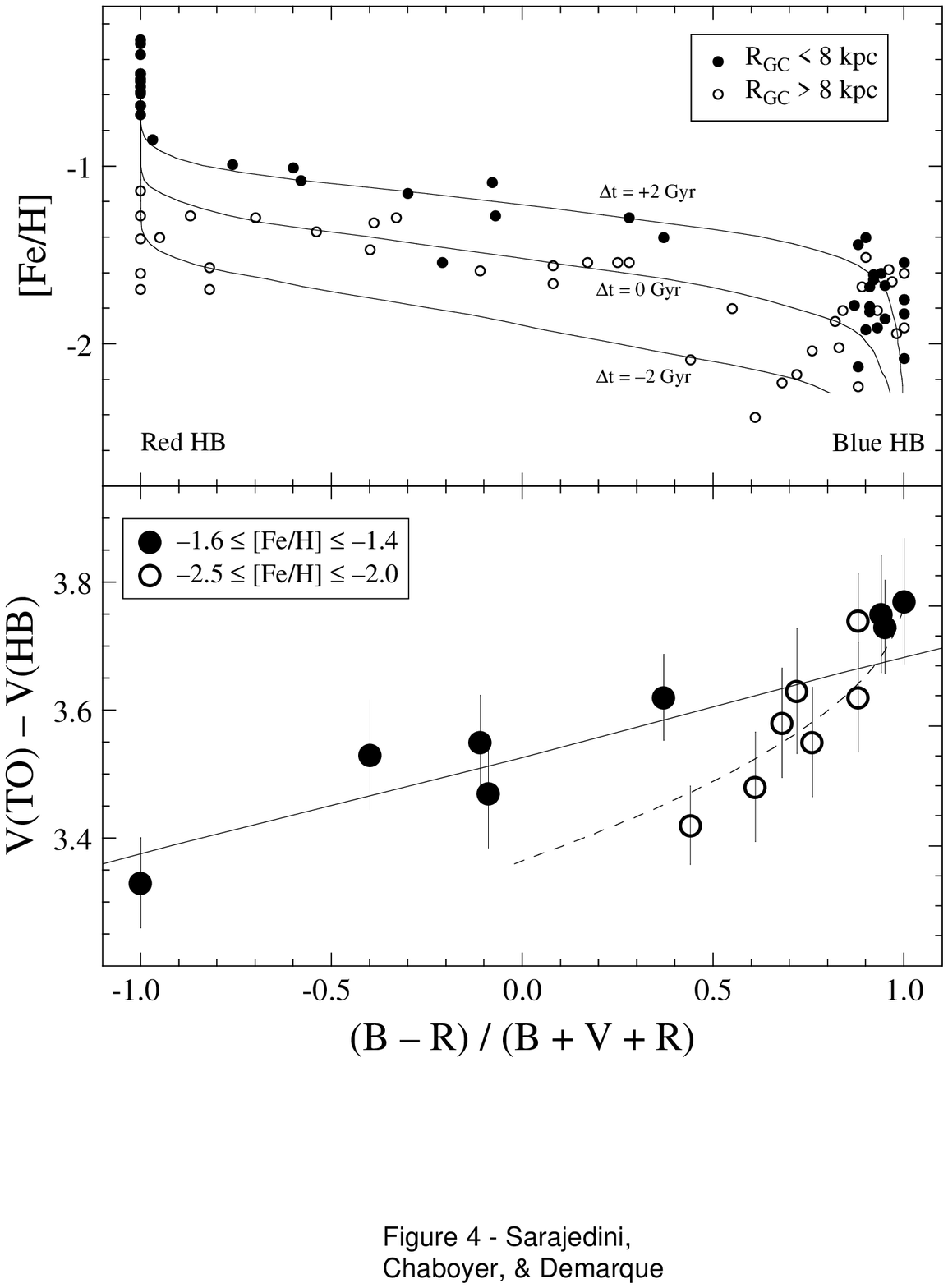]{The top panel illustrates the metal abundance of 
Milky Way globular clusters as a
function of their horizontal branch (HB) morphology as quantified by
$(B-R)/(B+V+R)$. This
represents the difference between the number of stars on the blue side
of the RR Lyrae instability strip (B) and on the red side (R), divided
by the total number of blue side, red side, and variable (V) stars on
the HB. Blue HB clusters have an index of $+1$ while red HB clusters
have an index of $-1$. Filled circles are clusters with Galactocentric
distances inside 8 kpc, while open circles are those outside 8 kpc.
The solid lines are the results of synthetic HB calculations (LDZ94) 
and indicate the expected locations of clusters for different relative 
ages. The lower panel shows the behavior of observed \dv\ values 
for clusters in two narrow ranges of metallicity. The solid and
dashed lines indicate the expected behavior based on the synthetic
models shown in the top panel.\label{fig4}}

\figcaption[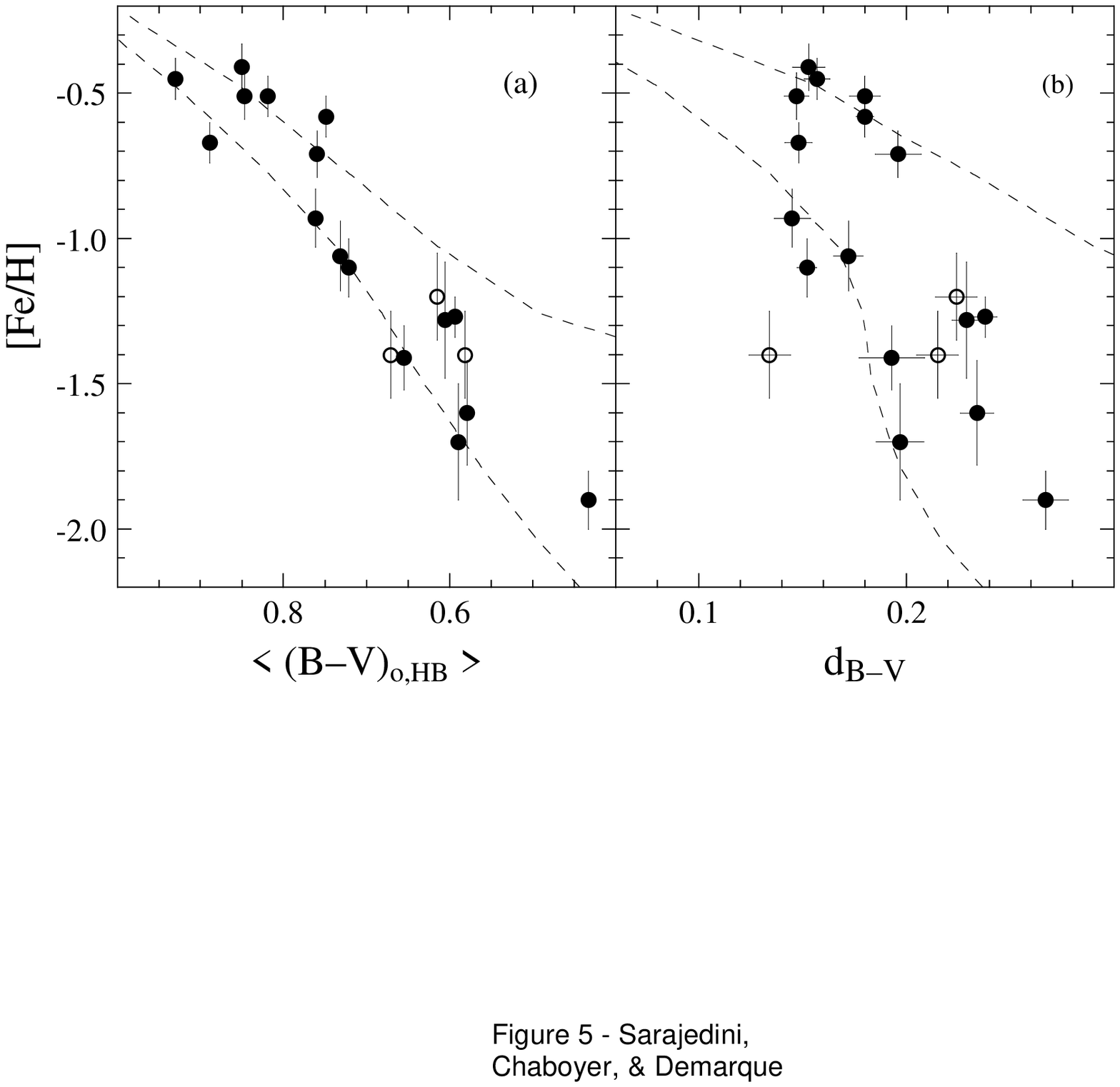]{(a) The observed values of metallicity and
mean dereddened horizontal branch (HB) color for globular clusters with
purely red HB morphologies. The open circles represent the clusters
Kron 3, NGC 121, and Pyxis. The first two are populous clusters in
the Small Magellanic Cloud and the latter one is a newly discovered
Milky Way globular cluster. The dashed lines represent the mean
intrinsic colors of zero age horizontal branches for scaled solar
abundances and masses of 0.90$M_{\odot}$ (left line) and 0.66$M_{\odot}$
(right line). (b) Same as in (a) except that the difference in color
between the HB and red giant branch is plotted. \label{fig5}}

\figcaption[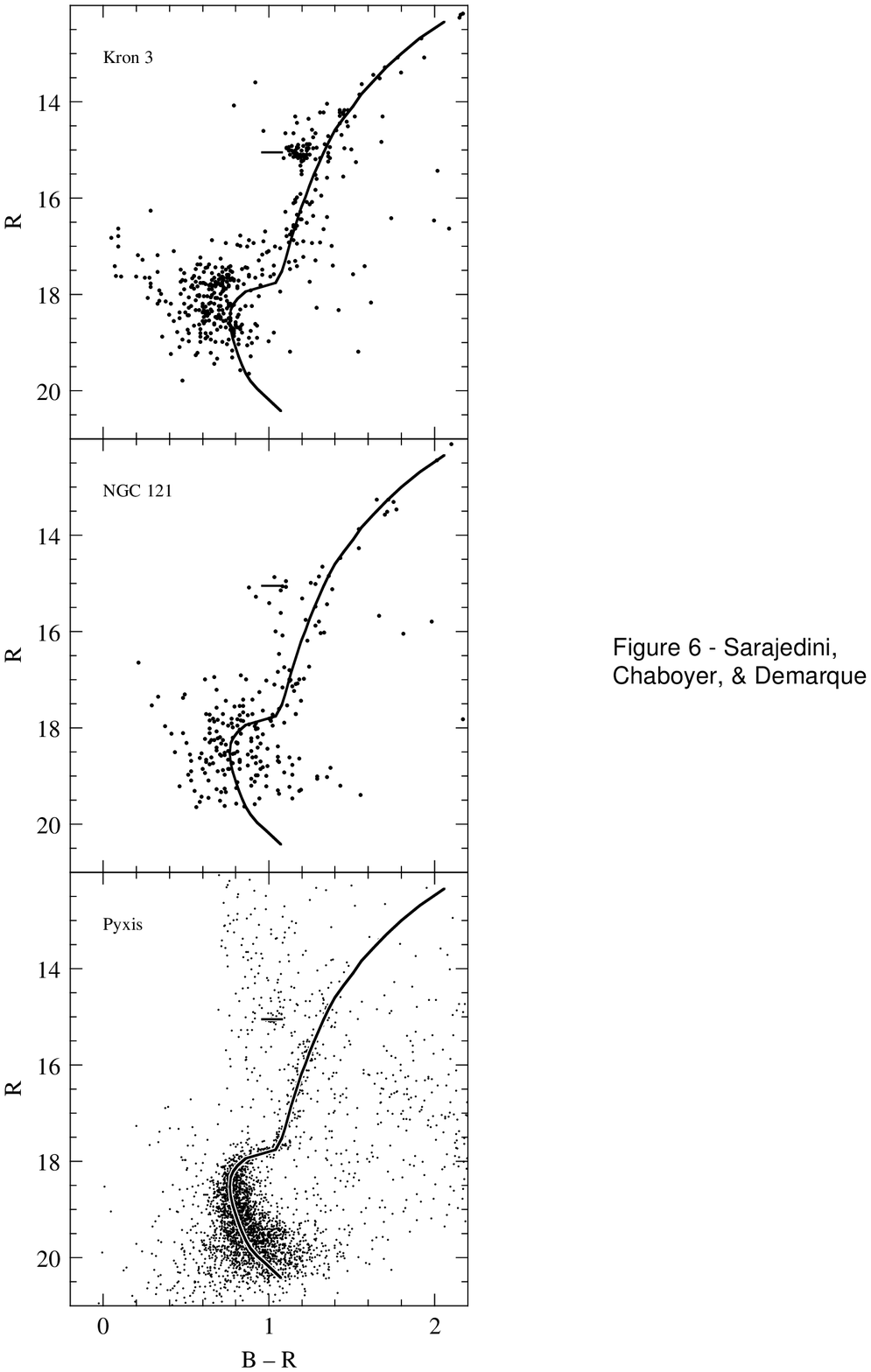]{Color-magnitude diagram comparisons between
Kron 3, NGC 121, Pyxis, and NGC 362. Each panel shows the $B-R$ photometry
for each of the first three clusters compared with the fiducial
sequence of NGC 362. Note that there is a correlation between the
relative main sequence turnoff locations and the relative colors of
the red horizontal branch. Relative to NGC 362, younger clusters 
(i.e. brighter turnoffs)
have redder horizontal branch colors.\label{fig6}}

\figcaption[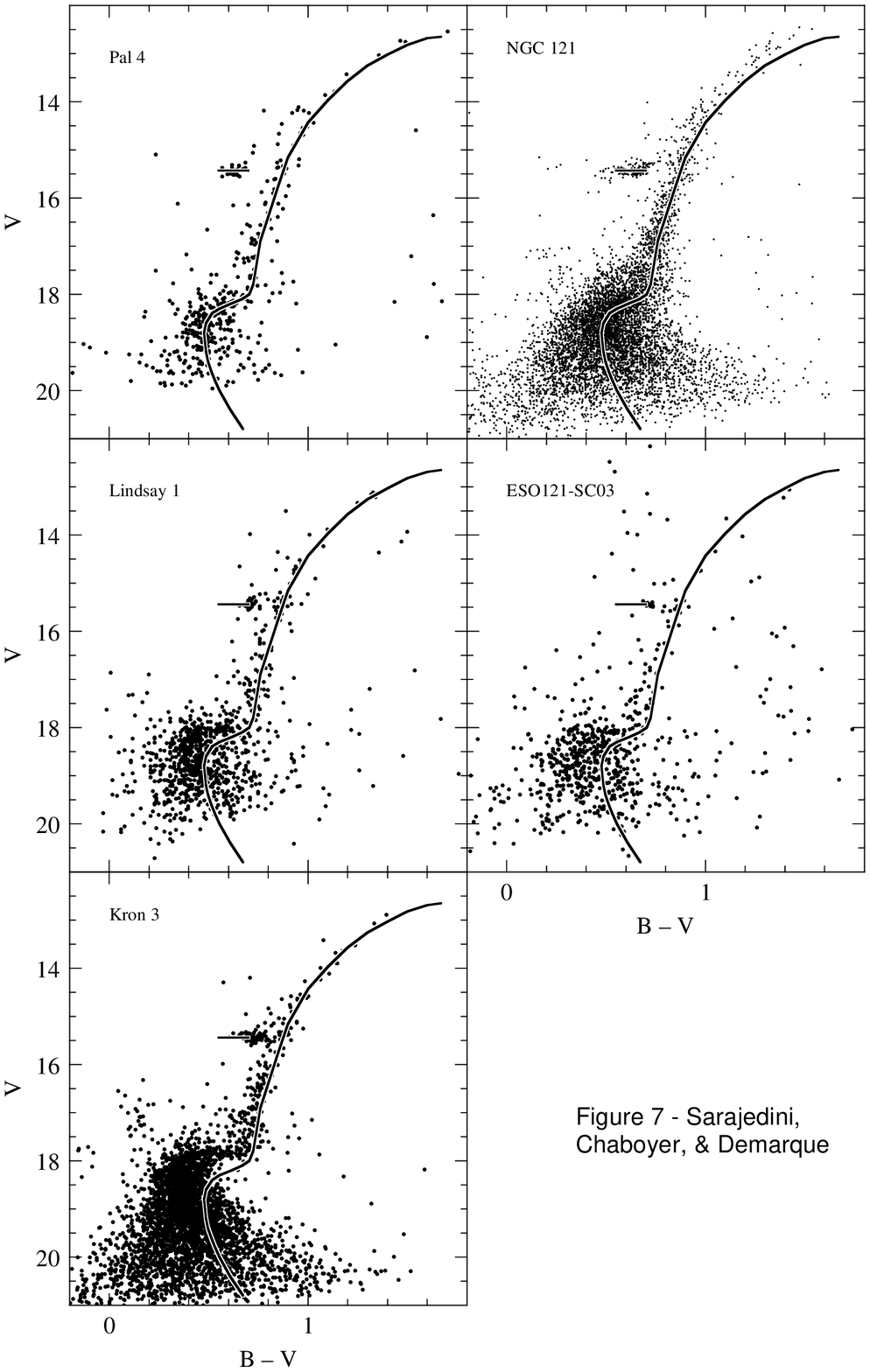]{Same as Fig. 6 except that $B-V$ photometry is
plotted for the globular clusters Palomar 4, NGC 121, Lindsay 1,
ESO121-SC03, and Kron 3 again compared with the fiducial sequence
of NGC 362. \label{fig7}}

\figcaption[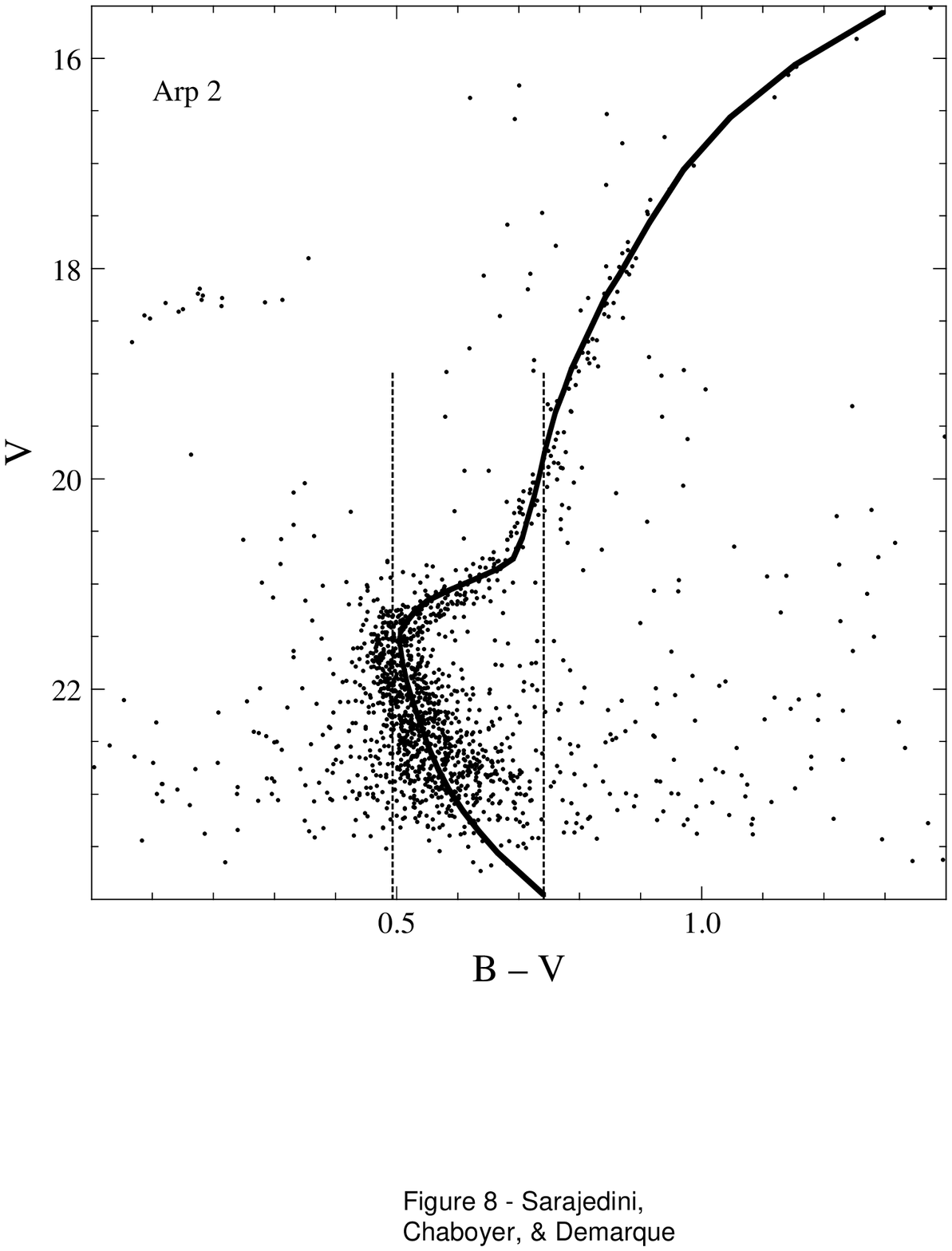]{The photometry for the globular cluster Arp 2
compared with the fiducial sequence of M68, where both have been
matched at the magnitude of the horizontal branch (HB) and the color
of the red giant branch at the level of the HB. The vertical dotted
lines represent the published value of \dbv\ for Arp 2. \label{fig8}}

\figcaption[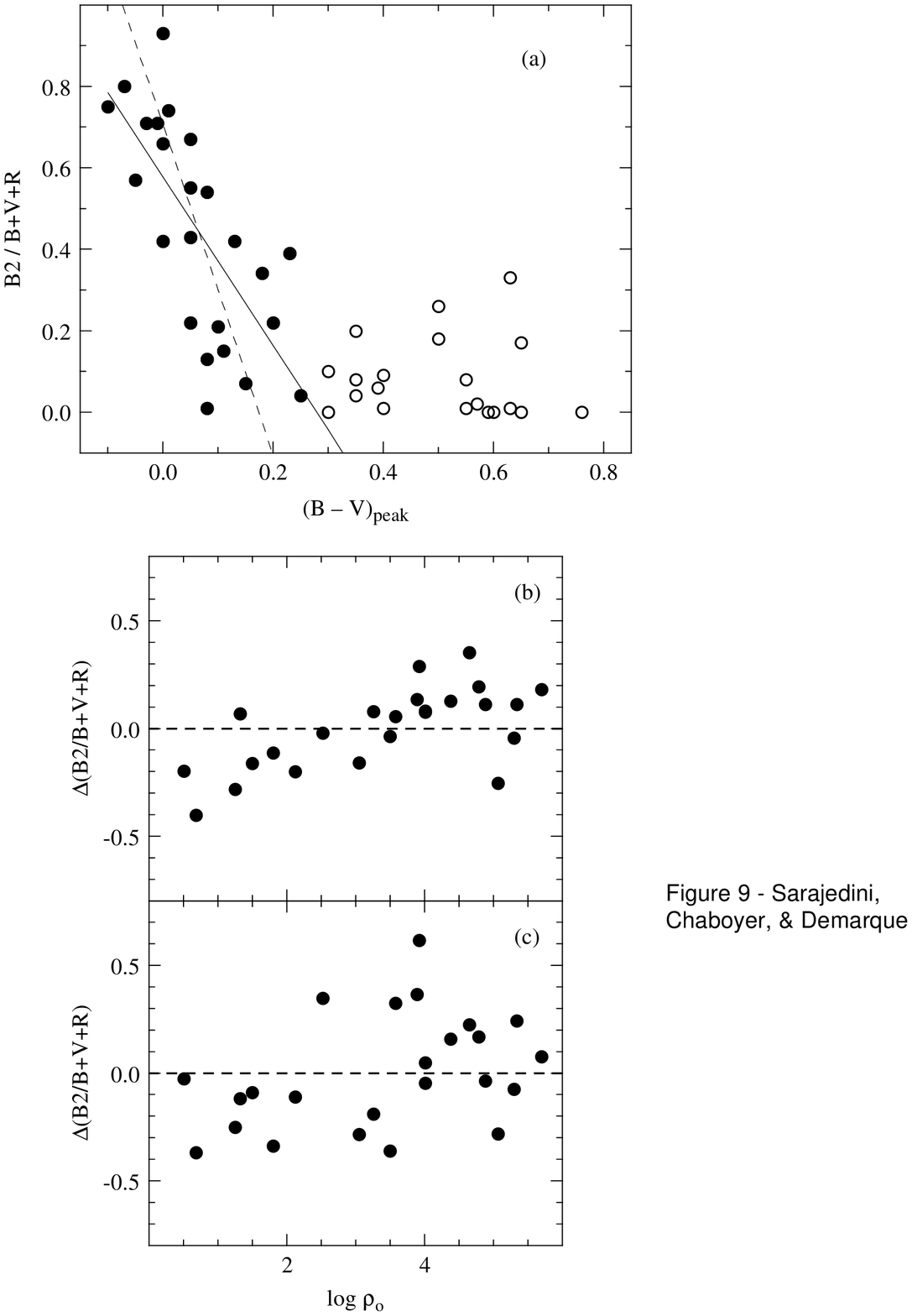]{(a) The horizontal branch (HB) morphology of Galactic
globulars as quantified by $B2/(B+V+R)$ as a function of the peak
dereddened color of the HB. The HB-type index is composed of
the number of stars blueward of $(B-V)_o = -0.02$ (i.e. the blue HB tail), 
denoted by B2, divided by the sum of the number of stars blueward of the 
instability strip (B), redward of the instability strip (R), and the number 
of RR Lyrae variables (V). The filled circles are the clusters 
to which the lines have been fit. The solid line 
represents the direct linear fit and the dashed line shows the `inverse' 
fit (i.e. independent and dependent variables are interchanged).
Figures (b) and (c) show the residuals from these fits, respectively, 
as a function of cluster central concentration. \label{fig9}}

\appendix
\clearpage
\section{Estimating the Gaussian Error in \dv}
In order to estimate the Gaussian error associated with the \dv\
measurements, CDS analyzed repeated observations.  However, as pointed
out by SVB, a few of the observations that CDS thought to be
independent could be traced back to the same original photometry.  For
this reason, we have repeated the original analysis of CDS, carefully
re-checking the literature to ensure that only independent \dv\
measurements are included.  We have discovered a few more errors in our
original work, and have searched the literature for new, independent
observations of \dv\ which have appeared since the work of
CDS.  The results of our literature search are presented in Table 1,
which gives the cluster name, \dv ~value and the
references.  Using the \dv ~values and errors, we have calculated and
tabulated the quantity $\delta \equiv (\dv_a - \dv_b)/(\epsilon_a^2 +
\epsilon_b^2)^{1/2}$, where $\dv_a$ is the measured \dv ~with its
error ($\epsilon_a$) as reported by observer $a$, and $\dv_b$ and
$\epsilon_b$ are the same quantities reported by observer $b$.
Essentially, $\delta$ is simply the difference in the \dv
~observations, normalized by the quoted errors.  If the observers are
quoting Gaussian 1-$\sigma$ error bars, then $\delta$ should have a
Gaussian distribution, with $\sigma = 1$.  There are 16 measurements
of $\delta$ in Table 1; for this sample size one would expect 5 values
of $\delta$ in excess of 1 for Gaussian errors.  However, this occurs
only once.  This suggests that the reported errors are an overestimate
of the Gaussian 1-$\sigma$ error bars.  Indeed, the F-test (Press \ea\
\cite{presea1992}) finds that there is only a 1\% chance that $\delta$
has a standard deviation of 1.0.  The quantity $\delta$ has an actual
standard deviation of 0.55.  If one choses the hypothesis that
$\delta$ has a standard deviation of 0.61 (as advocated by CDS in the
original analysis), this has a 69\% probability of being true from the
F-test.  Thus, multiplying the quoted errors by 0.61 yields a good
estimate of the Gaussian 1-$\sigma$ error in \dv.  This implies that
the original errors are actually $1.64\,\sigma$ error bars, which
corresponds to 90\% confidence limits.

\clearpage

\begin{table}[t]
  \begin{center}
  \begin{tabular}{lrrl}
\multicolumn{4}{c}{TABLE A.1}\\
\multicolumn{4}{c}{INDEPENDENT \dv ~OBSERVATIONS}\\[3pt]
\hline\hline
\multicolumn{1}{c}{Cluster}&
\multicolumn{1}{c}{\dv}&
\multicolumn{1}{c}{$\delta$}&
\multicolumn{1}{c}{References (TO,HB)}\\[2pt]
\hline
NGC 104& $3.61\pm 0.10$&&      	   Hesser \ea\ 1987\\
       & $3.81\pm 0.18$& $-0.971$&  Buonanno \ea\ 1989 \\
       & $3.62\pm 0.06$& $-0.086$&  Grundahl 1996 \\[3pt]
NGC 288& $3.73\pm 0.12$&&  Buonanno \ea\ 1989 \\
       & $3.70\pm 0.14$& $0.163$ & Pound \ea\ 1987; Olszewski \ea\ 1984 \\[3pt]
NGC 1261&    $3.57\pm 0.11$&&  Ferraro \ea\ 1993 \\
        &    $3.50\pm 0.14$& $0.393$&  Alcaino \ea\ 1992b \\[3pt]
NGC 1851&    $3.45\pm 0.10$&&  Walker 1992b \\
        &    $3.34\pm 0.10$& $0.778$&  Alcaino \ea\ 1990 \\[3pt]
NGC 3201&    $3.45\pm 0.21$&&   Brewer \ea\ 1993 \\
        &    $3.44\pm 0.12$& $0.041$& Alcaino \ea\ 1989; Cacciari 1984\\
	&    $3.40\pm 0.21$& $0.168$& Covino \& Ortolani 1997\\[3pt]
NGC 4590&    $3.42\pm 0.10$&&  Walker 1994 \\
        &    $3.49\pm 0.12$& $-0.448$ &   Buonanno \ea\ 1989\\
        &    $3.42\pm 0.10$& $0.000$ &  Alcaino \ea\ 1990; Harris 1975\\[3pt]
NGC 5897&    $3.60\pm 0.18$&&  Sarajedini 1992 \\
        &    $3.52\pm 0.18$& $0.314$&  Ferraro \ea\ 1992b \\[3pt]
NGC 5927&    $3.40\pm 0.14$&& Samus \ea\ 1996\\
	&    $3.51\pm 0.10$& $-0.639$ &Fullton \ea\ 1996\\
\hline
\end{tabular}
\end{center}
\label{taba2}
\end{table}
 
\begin{table}[t]
  \begin{center}
  \begin{tabular}{lrrl}
\multicolumn{4}{c}{TABLE A.1 (continued)}\\
\hline\hline
\multicolumn{1}{c}{Cluster}&
\multicolumn{1}{c}{\dv}&
\multicolumn{1}{c}{$\delta$}&
\multicolumn{1}{c}{References (TO,HB)}\\[2pt]
\hline
NGC 6121&    $3.68\pm 0.16$&&  Buonanno \ea\ 1989 \\
        &    $3.52\pm 0.10$& $0.848$&  Alcaino \ea\ 1988 \\
        &    $3.45\pm 0.13$& $1.116$&  Kanatas \ea\ 1995 \\[3pt]
NGC 6171&    $3.75\pm 0.18$&&  Buonanno \ea\ 1989 \\
        &    $3.70\pm 0.11$& $0.237$ &  Ferraro \ea\ 1991 \\[3pt]
NGC 6352&    $3.67\pm 0.10$&& Fullton \ea\ 1995\\
	&    $3.67\pm 0.06$& $0.000$ & Grundahl 1996\\[3pt]
Rup 106 &    $3.32\pm 0.07$&&  Buonanno \ea\ 1993 \\
        &    $3.27\pm 0.12$& $0.360$& Buonanno \ea\ 1990 \\
\hline
\end{tabular}
\end{center}
\label{taba1}
\end{table}

\clearpage

\plotone{fig1.eps}

\clearpage

\plotone{fig2.eps}

\clearpage

\plotone{fig3.eps}

\clearpage

\plotone{fig4.eps}

\clearpage

\plotone{fig5.eps}

\clearpage

\plotone{fig6.eps}

\clearpage

\plotone{fig7.eps}

\clearpage

\plotone{fig8.eps}

\clearpage

\plotone{fig9.eps}


\begin{thebibliography}{}

\bibitem[1996]{akbe1996}
Akritas, M. G., \& Bershady, M. A. 1996, \apj, 470, 706

\bibitem[1995]{bapiwa1995}
Bahcall, J. N., Pinsonneault, M. H. \& Wasserburg, G. J. 
1995, Rev. Mod. Phys., 67, 781

\bibitem[1995]{bapi1995}
Bailyn, C. D., \& Pinsonneault, M. H. 1995, \apj, 439, 705

\bibitem[1992]{bailea1992}
Bailyn, C. D., Sarajedini, A., Cohn, H., Lugger, P. M., \& Grindlay, J. E.
1992, \aj, 103, 1564

\bibitem[1993]{berg1993}
Bergbusch, P. A. 1993, \aj, 106, 1025

\bibitem[1992]{beva1992}
Bergbusch, P. A., \& VandenBerg, D. A 1992, \apjs, 81, 163

\bibitem[1989]{bol1989}
Bolte, M. 1989, \aj, 97, 1688

\bibitem[1992]{bol1992}
Bolte, M. 1992, \apjs, 82, 145

\bibitem[1994]{bol1994}
Bolte, M. 1994, \apj, 431, 223

\bibitem[1997]{borisea1997}
Borissova, J., Catelan, M., Spassova, N., \& Sweigart, A. 1997, \aj, 113, 692

\bibitem[1993]{buon1993}
Buonanno, R. 1993, in The Globular Cluster-Galaxy Connection,
ASP Conf. Ser. Vol. 48, edited by G. H. Smith \& J. P. Brodie 
(ASP:San Francisco), p. 131

\bibitem[1986]{buonea1986}
Buonanno, R., Buzzoni, A., Corsi, C. E. Fusi Pecci, F.,
Sandage, A. R. 1986, MSAIt, 57, 391

\bibitem[1994]{buonea1994}
Buonanno, R., Buzzoni, A., Corsi, C. E., Buzzoni, A., 
Cacciari, C., Ferraro, F. R. \& Fusi Pecci, F. 1994, \aap, 290, 69

\bibitem[1989]{bucofu1989}
Buonanno, R., Corsi, C. E., \& Fusi Pecci, F. 1989, \aap, 216, 80

\bibitem[1993]{buonea1993}
Buonanno, R., Corsi, C. E., Fusi Pecci, F., Richer, H. B.,
\& Fahlman, G. G. 1993, \aj, 105, 184

\bibitem[1995]{buonea1995}
Buonanno, R., Corsi, C. E., Fusi Pecci, F., Richer, H. B.,
\& Fahlman, G. G. 1995, \aj, 109, 650 (BCFRF)

\bibitem[1997]{buonea1997}
Buonanno, R., Corsi, C. E., Bellazzini, M., Ferraro, F. R.,
\& Fusi Pecci, F. 1997, \aj, 113, 706

\bibitem[1995]{buia1995}
Buonanno, R., \& Iannicola, G. 1995, in The Formation of the
Milky Way, edited by E. J. Alfaro \& A. J. Delgado 
(Cambridge University Press:Cambridge), p. 279

\bibitem[1991]{byle1991}
Byun, Y. -I., Lee, Y. -W. 1991, in The Formation and Evolution of Star
Clusters, ed. K. Janes (San Francisco:ASP), p. 243

\bibitem[1980]{car1980}
Carney, B. W. 1980, \apjs, 42, 481

\bibitem[1992a]{castjo1992a}
Carney, B. W., Storm, J., \& Jones, R. V. 1992a, \apj, 386, 663

\bibitem[1992b]{carnea1992b}
Carney, B. W., Storm, J., Trammel, S. R. \& Jones, R. V. 
1992b, \pasp, 104, 44

\bibitem[1995]{cafr1995}
Catelan, M. \& de Freitas Pacheco, J. A. 1995, \aap, 297, 345

\bibitem[1997]{chabea1997}
Chaboyer, B., Demarque, P., Kernan, P.J. \& Krauss, L.M. 1997, 
\apj, in press ({\tt astro-ph/9706128})

\bibitem[1996a]{chdesa1996a}
Chaboyer, B. C., Demarque, P., \& Sarajedini, A. 1996a, \apj, 459, 558

\bibitem[1996b]{chabea1996b}
Chaboyer, B. C., Demarque, P., Kernan, P. J., Krauss, L. M., 
\& Sarajedini, A. 1996b, \mnras, 283, 683

\bibitem[1992]{chsade1992}
Chaboyer, B. C., Sarajedini, A., \& Demarque, P. 1992, \apj, 394, 515

\bibitem[1986]{chhe1986}
Christian, C. A., \& Heasley, J. N. 1986, \apj, 303, 216

\bibitem[1985]{codede1985}
Cole, P. W., Demarque, P. \& Deupree, R. G. 1985, \apj, 291, 291

\bibitem[1990]{dca1990}
Da Costa, G. S., \& Armandroff, T. E. 1990, \aj, 100, 162

\bibitem[1996]{dcea1996}
D'Cruz, N., Dorman, B., Rood, R. T. \& O'Connell, R. W. 1996, \apj, 466, 359


\bibitem[1990]{dedeka1990}
Deliyannis, C. P., Demarque, P. \& Kawaler, S. D. 1990, \apjs, 73, 21

\bibitem[1960]{dem1960}
Demarque, P. 1960, \apj, 132, 366

\bibitem[1991]{dedesa1991}
Demarque, P. Deliyannis, C. P. \& Sarajedini, A. 1991, 
in Observational Tests of Cosmological Inflation, eds. T. Shanks, A. J. Banday, 
R. S. Ellis, C. S. Frenk \& A. W. Wolfendale, (Kluwer:Dordrecht), p.111

\bibitem[1997a]{deguki1997a}
Demarque, P., Guenther, D. B., \& Kim, Y. -C. 1997a, \apj, 474, 790

\bibitem[1988]{demea1988}
Demarque, P., Guenther, D. B., King, C. R. \& Green, E.M. 1988 
in Calibration of Stellar Ages, ed. A. G. D. Philip (L. Davis Press: 
Schenectady), p.101

\bibitem[1997b]{demea1997b}
Demarque, P., Lee, Y.-W., Yi, S., \& Zinn, R. 1997b, in preparation

\bibitem[1989]{demea1989}
Demarque, P., Lee, Y. -W., Zinn, R. J., \& Green, E. M. 1989,
in The Abundance Spread Within Globular Clusters: Spectroscopy of Individual
Stars, ed. G. Cayrel de Strobel, M. Spite, \& T. Lloyd Evans (Meudon, 
Observatoire de Paris), p. 97

\bibitem[1992]{dorm1992}
Dorman, B. 1992, \apjs, 80, 701

\bibitem[1991]{dicea1991}
Dickens, R. J., Croke, B. F. W., Cannon, R. D., \& Bell, R. A. 
1991, Nature, 351, 212 

\bibitem[1973]{egg1973}
Eggen, O. J. 1973, \apj, 182, 821

\bibitem[1962]{egsa1962}
Eggen, O. J. \& Sandage, A. R. 1962, \apj, 136, 735

\bibitem[1962]{eglysa1962}
Eggen, O. J., Lynden-Bell, D., \& Sandage, A. 1962, \apj, 136, 748 (ELS)

\bibitem[1966]{fau1966}
Faulkner, J. 1966, \apj, 144, 978

\bibitem[1966]{faib1966}
Faulkner, J., \& Iben, I. 1966, \apj, 144, 995

\bibitem[1997]{feas1997}
Feast, M. W. 1997, \mnras, 285, 339

\bibitem[1995]{ferrea1995}
Ferraro, I., Ferraro, F. R. Fusi Pecci, F., Corsi, C. E., \&
Buonanno, R. 1995, \mnras, 275, 1057

\bibitem[1993]{fuaxge1993}
Fuhrmann, K., Axer, M. \& Gehren, T. 1993, \aap, 271, 451

\bibitem[1996]{fusiea1996}
Fusi Pecci, F., Buonanno, R., Cacciari, C.,
Corsi, C. E., Djorgovski, S. G., Federici, L., Ferraro, F. R., 
Parmeggiani, G., \& Rich, R. M. 1996, \aj, 112, 1461

\bibitem[1993]{fusiea1993}
Fusi Pecci, F., Ferraro, F. R., Bellazzini, M., 
Djorgovski, S., Piotto, G., \& Buonanno, R. 1993, \aj, 105, 1145

\bibitem[1997]{gratea1997}
Gratton, R. G., Fusi Pecci, F., Carretta, E., Clementini, G., 
Corsi, C. E. \& Lattanzi, M.  1997, \apj, in press 
({\tt astro-ph/9704150})

\bibitem[1990]{grno1990}
Green, E. M., \& Norris, J. E. 1990, \apj, 353, L17

\bibitem[1996]{grun1996}
Grundahl, F. 1996, Ph.D. dissertation, University of Aarhus

\bibitem[1993]{gubrfu1993}
Guarnieri, M. D., Bragaglia, A., \& Fusi Pecci, F. 1993, \aaps, 102, 397

\bibitem[1989]{guen1989}
Guenther, D. B. 1989, \apj, 339, 1156

\bibitem[1997]{gude1997}
Guenther, D. B. \& Demarque, P. 1997, \apj, August 1 issue

\bibitem[1976]{har1976}
Hartwick, F.D.A. 1976, \apj, 209, 418

\bibitem[1982]{harr1982}
Harris, W. E. 1982, \apjs, 50, 573

\bibitem[1996]{harr1996}
Harris, W. E. 1996, http://www.physics.mcmaster.ca/Globular.html

\bibitem[1983]{ibre1983}
Iben, I. Jr. \& Renzini, A. 1983, \araa, 21, 271

\bibitem[1996]{jipa1996}
Jimenez, R., \& Padoan, P. 1996, \apj, 463, L17


\bibitem[1992]{jonea1992}
Jones, R. V., Carney, B. W., Storm, J., \& Latham, D. W. 1992, \apj, 386, 646

\bibitem[1996]{kal1996}
Kalyzny, J. 1996, preprint

\bibitem[1996]{koju1996}
Kov\'{a}cs, G., \& Jurcsik, J. 1996, \apj, 466, L17

\bibitem[1988]{kidegr1988}
King, C. R., Demarque, P., \& Green, E. M. 1988, in The 
Calibration of Stellar Ages, ed. A. G. D. Philip (Schenectady, Davis), p. 211

\bibitem[1995]{kjelea1995}
Kjeldsen, H., Bedding, T. R., Viskum, M. \& Frandsen, S. 1995, \aj, 109, 1313

\bibitem[1975]{lar1975}
Larson, R.B. 1975, \mnras, 173, 671

\bibitem[1990]{lar1990}
Larson, R.B. 1990, \pasp, 102, 79

\bibitem[1988]{ledezi1988}
Lee, Y. -W., Demarque, P., \& Zinn, R. J. 1988, in The Harlow
Shapley Symposium on Globular Cluster Systems in Galaxies, edited by
J. E. Grindlay \& A. G. D. Philip (Kluwer: Dordrecht), p. 505.

\bibitem[1991]{lee1991}
Lee, Y.-W. 1991 \apj, 373, L43

\bibitem[1990]{ledezi1990}
Lee, Y. -W., Demarque, P., \& Zinn, R. J. 1990, \apj, 350, 155 (LDZ)

\bibitem[1994]{ledezi1994}
Lee, Y. -W., Demarque, P., \& Zinn, R. J. 1994, \apj, 423, 248 (LDZ94)

\bibitem[1984]{mahosc1986}
Mateo, M., Hodge, P., \& Schommer, R. A. 1986, \apj, 311, 113

\bibitem[1981]{mesw1981}
Mengel, J. G. \& Sweigart, A. V. 1981,in IAU Coll. 68, 
Astrophysical Parameters for Globular Clusters, ed.A.G. D. Philip 
\& D. S. Hayes (Dordrecht:Reidel), 277

\bibitem[1976]{menogr1976}
Mengel, J. G., Norris, J. E. \& Gross, P. G. 1976, \apj, 204, 488

\bibitem[1997]{misafr1997}
Mighell, K. J., Sarajedini, A., \& French, R. 1997, in preparation

\bibitem[1980]{noar1980}
Noerdlinger, P. D. \& Arigo, R. J. 1980, \apj, 337, L15

\bibitem[1994]{norr1994}
Norris, J. E. 1994, \apj, 431, 645

\bibitem[1987]{olaasc1987}
Olszewski, E. W., Aaronson, M., \& Schommer, R. A. 1987, \aj, 93, 565

\bibitem[1975]{osth1975}
Ostriker, J.P. \& Thuan, T.X. 1975, \apj, 202, 353

\bibitem[1997]{pac1997}
Paczynski, B. 1997, in The Extragalactic Distance Scale, STScI 
May Institute, edited by M. Livio \& M. Donahue 
(Cambridge University Press:Cambridge), in press

\bibitem[1992]{presea1992}
Press, W. H., Teukolsky, S. A., Vetterling, W. T. \&
Flannery, B. P. 1992, Numerical Recipies in FORTRAN: The Art of 
Scientific Computing, Second Edition, (Cambridge University Press:Cambridge)

\bibitem[1977]{pime1977}
Pike, C. D., \& Meston, C. J. 1977, \mnras, 180, 613

\bibitem[1997]{pontea1997}
Pont, F. Mayor, M., Turon, C. \& VandenBerg, D. A. 1997, \aap, 
in press

\bibitem[1991]{prmi1991}
Proffit, C.R. \& Michaud, G. 1991, \apj, 371, 584

\bibitem[1997]{reid1997}
Reid, I. N. 1997, \aj, 114, 161

\bibitem[1991]{ren1991}
Renzini, A. 1991, in Observational Tests of Cosmological 
Inflation, edited by T. Shanks, A. J. Banday, R. S. Ellis, C. S. Frenk 
\& A. W. Wolfendale, (Dordrecht: Kluwer), p.131

\bibitem[1996]{renea1996}
Renzini, A. et al. 1996, \apj, 465, L23

\bibitem[1984]{ridamo1984}
Rich, R. M., Da Costa, G. S., \& Mould, J. R. 1984, \apj, 286, 517

\bibitem[1997]{richea1997}
Rich, R. M. et al. 1997, \apj, 484, L25

\bibitem[1995]{ricea1995}
Richer, H. et al. 1995, \apj, 451, L17

\bibitem[1996]{ricea1996}
Richer, H. et al. 1996, \apj, 463, 602

\bibitem[1987]{rith1987}
Ripley, B. D., \& Thompson, M. 1987, Analyst, 112, 377

\bibitem[1997]{ruba1997}
Rubenstein, E. P., \& Bailyn, C. D. 1997, \apj, 474, 701

\bibitem[1997]{sawe1997}
Salaris, M., \& Weiss, A. 1997, ({\tt astro-ph/9704238})

\bibitem[1986]{san1986}
Sandage, A. R. 1986, \araa, 24, 421

\bibitem[1981a]{san1981a}
Sandage, A. R. 1981a, \apj, 244, L23

\bibitem[1981b]{san1981b}
Sandage, A. R. 1981b, \apj, 248, 161 

\bibitem[1982]{san1982}
Sandage, A. R. 1982, \apj, 252, 553

\bibitem[1990]{san1990}
Sandage, A. R. 1990, \apj, 350, 603

\bibitem[1990]{saca1990}
Sandage, A., \& Cacciari, C. 1990, \apj, 350, 645

\bibitem[1959]{saeg1959}
Sandage, A. R. \& Eggen, O. J. 1959, \mnras, 119, 378

\bibitem[1967]{sawi1967}
Sandage, A. R., \& Wildey, R. 1967, \apj, 150, 469

\bibitem[1996]{sanea1996}
Sandquist, E. L., Bolte. M., Stetson, P. B., \& Hesser, J. E.
1996, \apj, 470, 910

\bibitem[1997]{saropi1997}
Saviane, I., Rosenberg, A., \& Piotto, G. 1997, preprint

\bibitem[1991]{sar1991}
Sarajedini, A. 1991, in Precision Photometry: Astrophysics of 
the Galaxy, edited by A. G. D. Philip, A. R. Upgren, \& K. A. Janes 
(Davis: Schenectady), p. 55

\bibitem[1993]{sara1993}
Sarajedini, A. 1993, \aj, 105, 2172

\bibitem[1990]{sade1990}
Sarajedini, A., \& Demarque, P. 1990, \apj, 365, 219

\bibitem[1996]{sage1996}
Sarajedini, A., \& Geisler, D. 1996, \aj, 112, 2013

\bibitem[1989]{saki1989}
Sarajedini, A., \& King, C. R. 1989, \aj, 98, 1624

\bibitem[1997]{sala1997}
Sarajedini, A., \& Layden, A. C. 1997, \aj, 113, 264 (SL97)

\bibitem[1995]{salele1995}
Sarajedini, A., Lee, Y. -W., \& Lee, D. -H. 1995, \apj, 450, 712 (SLL)

\bibitem[1978]{sezi1978}
Searle, L. \& Zinn, R. J. 1978, \apj, 225, 357 (SZ)

\bibitem[1994]{silbea1994}
Silbermann, N. A., Smith, H. A., Bolte. M., \& Hazen, M. L.
1994, \aj, 107, 1764

\bibitem[1993]{skilea1993}
Skillen, I., Fernley, J. A., Stobie, R. S., \& Jameson, R. F. 1993, 
\mnras, 265, 301

\bibitem[1997]{sosea1997}
Sosin, C. et al. 1997, \apj, 480, L35

\bibitem[1989]{stetea1989}
Stetson, P. B., Hesser, J. E., Smith, G. H., VandenBerg, D. A., \& Bolte, M.
1989, \aj, 97, 1360

\bibitem[1996]{stvabo1996}
Stetson, P. B., VandenBerg, D. A., \& Bolte, M., 1996, PASP, 108, 560 (SVB)

\bibitem[1985]{stdamo1985}
Stryker, L. L., Da Costa, G. S., \& Mould, J. R. 1985, \apj, 298, 544

\bibitem[1987]{swei1987}
Sweigart, A. V. 1987, \apjs, 65, 95

\bibitem[1994]{swei1994}
Sweigart, A. V. 1994, \apj, 426, 612

\bibitem[1975]{tin1975}
Tinsley, B. M. 1975, \apj, 197, 159

\bibitem[1967]{van1967}
van den Bergh, S. 1967, \aj, 72, 70 

\bibitem[1993]{vamo1993}
van den Bergh, S., \& Morris, S. 1993, \aj, 106, 1853

\bibitem[1990]{vadu1990}
VandenBerg, D. A., \& Durrell, P. R. 1990, \aj, 99, 221

\bibitem[1990]{vabost1990}
VandenBerg, D. A., Bolte, M., \& Stetson, P. B. 1990, \aj, 100, 445

\bibitem[1996]{vame1996}
van't Veer-Menneret, C. \& M\'{e}gessier, C. 1996, \aap, 309, 879

\bibitem[1992]{walk1992}
Walker, A. 1992, \pasp, 100, 1063

\bibitem[1994]{walk1994}
Walker, A. 1994, \aj, 108, 555

\bibitem[1996]{wane1996}
Walker, A. R., \& Nemec, J. M. 1996, \aj, 112, 2026

\bibitem[1962]{wildea1962}
Wildey, R. L., Burbidge, E. M., Sandage, A. R. \& Burbidge, G. R. 1962, 
\apj, 135, 94

\bibitem[1993]{yilede1993}
Yi, S., Lee, Y., -W. \& Demarque, P. 1993, \apj, 411, L25

\bibitem[1985]{zinn1985}
Zinn, R. J. 1985, \apj, 293, 424

\bibitem[1986]{zinn1986}
Zinn, R. J. 1986, in Stellar Populations (Cambridge:Cambridge University
Press), p. 73

\bibitem[1993]{zinn1993}
Zinn, R. J. 1993, in The Globular Cluster-Galaxy Connection,
ed. G. H. Smith \& J. P. Brodie (San Francisco:ASP), p. 38

\bibitem[1984]{ziwe1984}
Zinn, R. J., \& West, M. J. 1984, \apjs, 55, 45

\end{thebibliography}
\end{document}